\documentclass[journal]{IEEEtran}

\usepackage{amsmath}
\usepackage{subfigure}
\usepackage{lineno,hyperref}
\usepackage{xcolor}
\usepackage{extarrows}
\usepackage{algorithm}
\usepackage{algorithmic}
\usepackage{setspace}
\usepackage{amssymb}
\usepackage{bm}
\usepackage{graphicx}
\usepackage[justification=centering]{caption}

\newcommand{\ba}{\boldsymbol{a}}
\newcommand{\bb}{\boldsymbol{b}}
\newcommand{\bc}{\boldsymbol{c}}
\newcommand{\bd}{\boldsymbol{d}}
\newcommand{\bx}{\boldsymbol{x}}
\newcommand{\bone}{\boldsymbol{1}}

\newcommand{\by}{\boldsymbol{y}}

\newcommand{\bz}{\boldsymbol{z}}
\newcommand{\bp}{\boldsymbol{p}}
\newcommand{\br}{\boldsymbol{r}}
\newcommand{\bs}{\boldsymbol{s}}
\newcommand{\bq}{\boldsymbol{q}}

\newcommand{\blambda}{\boldsymbol{\lambda}}

\newcommand{\bX}{\boldsymbol{X}}
\newcommand{\bS}{\boldsymbol{S}}

\newcommand{\bA}{\boldsymbol{A}}

\newcommand{\bI}{\boldsymbol{I}}
\newcommand{\bF}{\boldsymbol{F}}
\newcommand{\bC}{\boldsymbol{C}}

\newcommand{\bU}{\boldsymbol{U}}
\newcommand{\bV}{\boldsymbol{V}}
\newcommand{\bH}{\boldsymbol{H}}

\newcommand{\bLambda}{\boldsymbol{\Lambda}}

\newcommand{\N}{\mathcal{N}}

\newcommand{\mO}{\mathcal{O}}
\newcommand{\mA}{\mathcal{A}}

\newcommand{\bnu}{\boldsymbol{\nu}}

\newcommand{\bPhi}{\boldsymbol{\Phi}}

\newcommand{\bomega}{\boldsymbol{\omega}}

\begin{document}
\title{Iterative Detection for Orthogonal Time Frequency Space Modulation {with Unitary Approximate Message Passing}}
\author{Zhengdao~Yuan, Fei Liu, Weijie Yuan, \IEEEmembership{ Member,~IEEE,} Qinghua Guo, \IEEEmembership{Senior Member,~IEEE}, Zhongyong Wang, and Jinhong Yuan, ~\IEEEmembership{Fellow, IEEE}
\thanks{The work of Z. Yuan, F. Liu and Z. Wang was supported by the Postdoctoral science foundation of China (2019M652576), Science and technology research project of Henan (202102210313), Henan research project of high education (20B510005) and National Natural Science Foundation of China (NSFC61901417).}
\thanks{Z. Yuan is with the Artificial Intelligence Technology Engineering Research Center, Henan Radio \& TV University, and School of Information Engineering, and also with Zhengzhou University, Zhengzhou 450002, China (e-mail: yuan\_zhengdao@163.com).}
\thanks{F. Liu and Z. Wang is with the School of Information Engineering, Zhengzhou University, Zhengzhou 450002, China, (e-mail: ieliufei@hotmail.com, zywangzzu@gmail.com)}
\thanks{W. Yuan and J. Yuan are with the School of Electrical Engineering and Telecommunications, University of New South Wales, NSW 2052, Australia (e-mail: weijie.yuan@unsw.edu.au, j.yuan@unsw.edu.au).}
\thanks{Q. Guo is with the School of Electrical, Computer and Telecommunications Engineering, University of Wollongong, Wollongong, NSW 2522, Australia  (e-mail: qguo@uow.edu.au).}
}
\maketitle

\begin{abstract}

The orthogonal-time-frequency-space (OTFS) modulation has emerged as a promising modulation scheme for
high mobility wireless communications. To harvest the time and frequency diversity promised by OTFS, some promising detectors, especially message passing based ones, have been developed by taking advantage of the sparsity of the channel in the delay-Doppler domain. However, when the number of channel paths is relatively large or fractional Doppler {shifts have} to be considered, the complexity of existing detectors is a concern, and the message passing based detectors may suffer from performance loss due to the short loops involved in message
passing. In this work, we investigate the design of OTFS detectors based on the approximate message passing (AMP). In particular, {leveraging the unitary AMP (UAMP), we design new detectors that enjoy} the structure of the channel matrix and allow efficient implementation. In addition, the estimation of noise variance is incorporated into the UAMP-based detectors. Thanks to the robustness of UAMP relative to AMP, the UAMP-based detectors deliver superior performance, and outperform state-of-the-art detectors significantly. {We also investigate iterative} joint detection and decoding in a coded OTFS system, where the OTFS detectors are integrated into a powerful turbo receiver, leading to considerable performance gains.

\end{abstract}


\section{Introduction}
Recently the orthogonal time frequency space (OTFS) modulation has attracted much attention due to its capability of achieving reliable communications for high mobility applications \cite{Hadani2017,Raviteja2018,Surabhi2019, OAP, shuangyang2020twc}. OTFS offers both time and frequency diversity as each symbol is spread over the time and frequency domains through the two dimensional (2D) inverse symplectic finite Fourier transform (ISFFT) \cite{Hadani2017,Raviteja2018}. Compared with orthogonal frequency division multiplexing (OFDM), OTFS can achieve significant performance gains in high mobility scenarios \cite{Farhang2018}. In addition, when the number of channel paths is small, the effective channel in the delay-Doppler (DD) domain is sparse, which allows efficient channel estimation and data detection using message passing techniques \cite{Raviteja2018}.

A practical and powerful detector is essential to harvest the full time and frequency diversity promised by OTFS. The optimal maximum \emph{a posteriori} (MAP) detector is impractical due to its complexity growing exponentially with the length of the OTFS block. In \cite{LiA2017}, an effective  channel matrix in the DD domain was derived, based on which a low-complexity two-stage detector was proposed. The authors in \cite{zemen2017} proposed a detection scheme, which uses minimum mean squared error (MMSE) equalization in the first iteration, followed by parallel interference cancellation with a soft-output sphere decoder in subsequent iterations. {A low complexity iterative rake decision feedback detector was proposed in \cite{rakea}, \cite{rake}, which extracts and coherently combines the multiple copies of the symbols (due to multipath propagation) in the DD grid using maximal ratio combining (MRC). Multiple modified MRC detectors with enhanced performance were also developed in \cite{rake}. They performs similarly or even better than the message passing based detector \cite{Raviteja2018} while with a linear complexity.} The design of low-complexity detectors were also investigated based on the message passing techniques \cite{Raviteja2018}, \cite{Raviteja2019}, \cite{Raviteja2017}.
A variational Bayes (VB) based detector was proposed in \cite{Yuan2019} to achieve better convergence compared with the existing message passing based detectors. The detectors in \cite{Raviteja2018}, \cite{Yuan2019}, and \cite{otfslmmserecv} take advantage of the sparsity of the channel matrix in the DD domain, and their complexity depends on the number of nonzero elements in each row of the channel matrix, which is denoted by $S$. Without considering fractional Doppler shifts, $S$ is equal to the number of channel paths. In general, a wideband system is able to provide sufficient delay resolution, however, the Doppler resolution depends on the time duration of the OTFS block. To fulfill the low latency requirement in future wireless communications, the time duration should be relatively small, where it is necessary to consider fractional Doppler shifts \cite{Raviteja2018,Raviteja2019}. In this case, the value of $S$ is proportional to the number of paths and can be significantly larger than the number of channel paths. {This} leads to two issues: the complexity of the existing detectors is a concern especially in the case of rich scattering environments, and the message passing based detectors may suffer from performance loss due to the presence of short loops in the corresponding system graph model.


This work aims to address the above issues by designing efficient detectors based on the approximate message passing (AMP) algorithm \cite{Donoho2010a}, \cite{Donoho2010b}. The use of AMP in this work is due to two facts: AMP was developed based on loopy belief propagation in dense factor graphs, i.e., short loops can be handled by AMP, and AMP has a low complexity.  It is known that AMP works well for i.i.d. (sub-)Gaussian system transfer matrix, but it may suffer from performance loss or even diverge for a general system transfer matrix \cite{Caltagirone2014}. Inspired by \cite{Guo2013}, the work in \cite{Guo2015UtAMP} showed that AMP can still work well for a general system transfer matrix when a unitary transform of the original model is used, where
the unitary matrix for transformation can be the conjugate transpose of the left singular matrix of the general system transfer matrix \cite{Guo2015UtAMP, BiUTAMP} obtained through singular value decomposition (SVD).
{This variant to AMP is called unitary AMP (UAMP), which is formerly known as UTAMP. UAMP} has been {used for low complexity robust} sparse Bayesian learning \cite{UTAMPSBL},  bilinear recovery \cite{BiUTAMP}, inverse synthetic aperture radar (ISAR) imaging with high Doppler resolution \cite{ISAR}, direction of arrival (DOA) estimation \cite{DOAUT}, etc. It will be shown in this paper that UAMP is well suitable for OTFS because the channel matrix in the DD domain is a block circulant matrix with circulant block (BCCB), {and the 2D discrete Fourier transform can be used for the unitary transformation, thereby allowing an efficient implementation} using the 2D fast Fourier transform (FFT) algorithm. This leads to an attractive UAMP-based detector in two aspects. First, the complexity of the UAMP-based detector is in the order of the logarithm of the OTFS block length per symbol {per iteration}, which is independent of $S$. Second, thanks to the robustness of UAMP, the UAMP-based detector delivers significantly better performance. In addition, as the noise variance is normally unknown in practice, the noise variance estimation is incorporated in the design of the UAMP-based detector in this work, {so that an extra noise variance estimator is not required (existing OTFS detectors except the one in \cite{rakea} often assume a perfect knowledge of the noise variance)}. It is shown that the UAMP-based detector can significantly outperform the AMP-based detector and other state-of-the-art detectors \cite{Raviteja2018}, \cite{rakea, rake}, \cite{Yuan2019}. In this paper, we also extend our investigations to coded OTFS systems, and integrate (U)AMP-based detectors into a powerful turbo receiver to achieve joint iterative detection and decoding.


The remainder of the paper is organized as follows. In Section II, the OTFS system model is introduced. We design (U)AMP-based detectors with bi-orthogonal waveform in Section III, and rectangular waveform in Section IV. {The discussion is extended to iterative joint detection and decoding in a coded OFTS system and the performance prediction of the OTFS system with the proposed detectors is discussed in Section V.} Simulations results are provided in Section VI, followed by conclusions in Section VII.


\textit{Notations}- Boldface lower-case and upper-case letters denote vectors and matrices, respectively. The superscripts $(\cdot)^H$ and $(\cdot)^{T}$ represent conjugate transpose and transpose operations, respectively. The superscript $*$ is used to denote the conjugate operation. We define $[\cdot]_M$ as the mod $M$ operation. The probability density function of a complex Gaussian variable with mean {$\hat x$} and variance $\nu_x$ is represented by $\N(x|{\hat x},\nu_x)$. The notation $\left\langle f(\mathbf{x}) \right\rangle _{q(\mathbf{x})}$ denotes the expectation of the function $f(\mathbf{x})$ with respect to probability density $q(\mathbf{x})$. The relation $f(x)=cg(x)$ for some positive constant $c$ is written as $f(x)\propto g(x)$. Notation $\otimes$ represents the Kronecker product, and $\ba\cdot\bb$ and $\ba\cdot/\bb$ represent the component-wise product and division between vectors $\ba$ and $\bb$, respectively. The notation $\bX=\text{reshape}_M\left(\bx\right)$ represents that the vector $\bx$ is reshaped as an $M \times N$ matrix $\bX$ column by column, where the length of vector $\bx$ is $MN$. We use $\bx=\text{vec}\left(\bX\right)$ to represent vectorization of matrix $\bX$ column by column. The notation $\text{diag}(\ba)$ represents a diagonal matrix with the elements of $\ba$ as its diagonal. We use $|\bA|^2$ to denote the element-wise magnitude squared operation for matrix $\bA$. We use $\textbf{1}$ and $\textbf{0}$ to denote an {all-ones} vector and an all-zeros vector with a proper length, respectively. We use $q_j$ to denoted the $j$th entry of $\bq$.  The superscript $t$ of $\textbf{s}^t$ denotes the iteration index in an iterative algorithm.

\begin{figure*}[htbp]
	\centering
	\includegraphics[width=1.8\columnwidth]{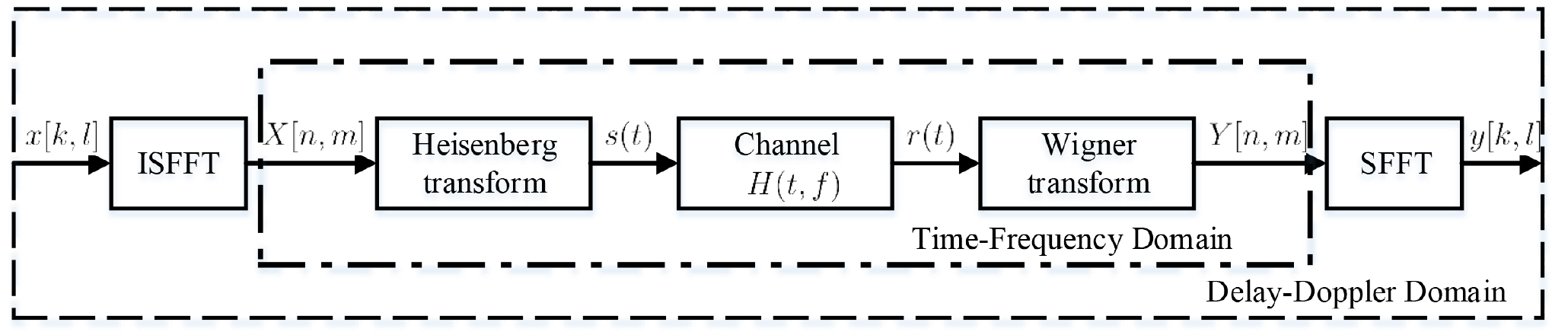}
	\caption{OTFS modulation and demodulation \cite{Raviteja2018}. }
	\label{fig:model}
\end{figure*}

\section{System Model}
The OTFS modulation and demodulation are shown in Fig. \ref{fig:model}, which are implemented with 2D inverse SFFT (ISFFT) and SFFT at the transmitter and receiver, respectively \cite{Hadani2017}, \cite{Monk2016OTFSO}.
{A (coded) bit sequence is mapped to symbols $\{x[k,l], k = 0,\cdots,N-1, l =0,\cdots M-1\}$ in the DD domain, where $x[k,l]\in\mA=\{\alpha_1, ....\alpha_{|\mA|}\}$ with $|\mA|$ being the cardinality of $\mA$, $l$ and $k$ denote the indices of delay and Doppler shifts, respectively, and $N$ and $M$ are the number of grids of the DD plane.  As shown in Fig. \ref{fig:model}, ISFFT is performed to convert the symbols to signals in the time-frequency (TF) domain, i.e.,
\begin{eqnarray}
X_{tf}[n,m] = \frac{1}{\sqrt{MN}}\sum_{k=0}^{N-1}\sum_{l=0}^{M-1}x[k,l]e^{j2\pi(\frac{nk}{N}-\frac{ml}{M})}.
\end{eqnarray}
Then the signals $\{X_{tf}[m,n]\}$ in the TF domain are converted to a continuous-time waveform $s(t)$ using the Heisenberg transform with a transmit waveform $g_{tx}(t)$, i.e.,
\begin{align}
s(t) = \sum_{n=0}^{N-1}\sum_{m=0}^{M-1}X_{tf}[n, m]g_{tx}(t-nT)e^{j2\pi m\Delta f(t-nT)},
\end{align}
where $\Delta f$ is subcarrier spacing and $T = 1/\Delta f$.

The signal $s(t)$ is then transmitted through a time-varying channel and the received signal in the time domain can be expressed as
\begin{align}
u(t) = \int\int h(\tau,\nu)s(t-\tau)e^{j2\pi\nu(t-\tau)} d\tau d\nu,
\end{align}
with $h(\tau, \nu)$ being the channel impulse response in the (continuous) DD domain. The channel impulse response can be expressed as
\begin{align}
h(\tau, \nu) = \sum_{i=1}^P h_i\delta(\tau-\tau_i)\delta(\nu - \nu_i),
\end{align}
where $\delta(\cdot)$ is the Dirac delta function, $P$ is the number of resolvable propagation paths, and $h_i$, $\tau_i$ and $\nu_i$ represent the gain, delay and Doppler shift associated with the $i$th path, respectively. The delay and Doppler-shift taps for the $i$th path are given by
\begin{eqnarray}
\tau_i=\frac{l_i}{M\Delta f}, \\
\nu_i=\frac{k_i+\kappa_i}{NT}, \label{eq:ddtap}
\end{eqnarray}
where $l_i$ and $k_i$ are the delay index and Doppler index of the $i$th path, and $\kappa_{{i}}$ $\in[-\frac{1}{2},\frac{1}{2}]$ is a fractional Doppler associated with the $i$th path. Here, $M\Delta f$ is the total bandwidth of the system and $NT$ is the duration of an OTFS block.

At the receiver side, a receive waveform $g_{rx}(t)$ is used to transform the received signal $r(t)$ to the TF domain i.e.,
\begin{align}
Y(t,f) = \int g_{rx}^{*}(t'-t)u(t')e^{-j2\pi f(t'-t)}dt',
\end{align}
which is then sampled at $t=nT$ and $f=m\Delta f$, yielding $Y[n, m]$.
Finally, the SFFT is applied to $\{Y[n, m]\}$ to obtain the signal $y[k,l]$ in the DD domain, i.e.,
\begin{align}
y[k,l] = \frac{1}{\sqrt{MN}} \sum_{n=0}^{N-1}\sum_{m=0}^{M-1}Y[n,m]e^{-j2\pi(\frac{nk}{N} - \frac{ml}{M})}.
\end{align}
}

\section{(U)AMP-Based Detection for OTFS with Bi-orthogonal Waveform}
\subsection{System Model in the Delay Doppler Domain}
Assume that the transmit waveform and receive waveform satisfy the bi-orthogonal property \cite{Hadani2017}, then the channel input-output relationship in the DD domain can be written as \cite{Raviteja2018,Raviteja2019Practical}
\begin{eqnarray}
&&y[k,l]=\sum_{i=0}^{P-1}\sum_{c=-N_i}^{N_i}h_i x([k-k_i+c]_N,[l-l_i]_M)\nonumber \\
&& \ \ \ \ \ \ \times\frac{1}{N}\frac{1-e^{-j2\pi(-c-\kappa_i)}}{1-e^{-j2\pi\frac{-c-\kappa_i}{N}}}e^{-j2\pi \frac{l_i(k_i+\kappa_i)}{MN}}
  + \omega[k,l], \nonumber \\ \label{eq:IdealFracY}
\end{eqnarray}
where
$N_i<N$ is an integer, and $ \omega[k,l]$ is the noise in the DD domain. We can see that for each path, the transmitted signal is circularly shifted, and scaled by a corresponding channel gain. We reshape $x[k,l]$ into a vector $\bx \in \mathbb{C}^{MN\times 1}$, where the $j$th element $x_j$ is $x[k,l]$ with  $j=kM+l$. Similarly, a vector $\by\in\mathbb{C}^{MN\times 1}$ can also be constructed based on $\{{y}[k,l]\}$. Then the channel input-output relationship in \eqref{eq:IdealFracY} can be rewritten in a vector form as
\begin{eqnarray}
\by=\bH\bx+\bomega, \label{eq:IdealFracY2}
\end{eqnarray}
where $\bH\in\mathbb{C}^{MN\times MN}$ is the effective channel in the DD domain, and $\bomega$ denotes a white Gaussian noise with mean 0 and variance $\epsilon^{-1}$ (or precision $\epsilon$).

\subsection {AMP-Based Detector}

Based on \eqref{eq:IdealFracY2}, we aim to recover the discrete-valued symbols in the vector $\bx$. Motivated by the capability of dealing with short loops and the low complexity of AMP, we estimate $\bx$ using the AMP algorithm \cite{Donoho2010a}, \cite{Rangan2011}. The AMP-based detection algorithm is derived in the following and summarized in Algorithm \ref{AMPv1}.

The AMP algorithm decouples the estimation of vector $\bx$.  Lines 1-6 of Algorithm \ref{AMPv1} follow the AMP algorithm directly, and $\bq$ computed in Line 6 can be regarded as pseudo observations with the following decoupled pseudo observation model
\begin{equation}
q_j
=
x_j + \varpi_j,~ j=1,..., MN,
\label{q_n}
\end{equation}
where $\varpi_j $ denotes a Gaussian noise with mean zero and variance $\nu_{q_j}$, which is computed in Line 5 of the algorithm.

Lines 7-10 are used to compute the \emph{a posteriori} mean and variance of each $x_j$ based on the above pseudo observation model and the prior of $x_j$, which is a uniform discrete distribution, i.e.,
\begin{equation}
P(x_j=\alpha_a) ={1}/{|\mA|}.
\end{equation}
It is not hard to show that the \emph{a posteriori} mean $\hat x_j$ and variance $\nu_{x_j}$ of $x_j$ are given by

\begin{algorithm}
	\setstretch{1.25}
	\caption{AMP-based OTFS Detector}
	Initialize $\bs^{(-1)}=\boldsymbol{0}$, $\hat\bx^{(0)}=\boldsymbol{0}$, $\bnu_x^{(0)}=\boldsymbol{1}$, and $t=0$. \\
	\textbf{Repeat}
	\begin{algorithmic}[1]
		\STATE $\bnu_{\bp}=\left|\bH\right|^2\bnu_{\bx}^t$\\
		\STATE $\bp=\bH\hat\bx^{t}-\bnu_{\bp}\cdot\bs^{t-1}$\\
		\STATE $\bnu_{\bs}=\textbf{1}./\left(\bnu_{\bp}+\epsilon^{-1} \textbf{1}\right)$\\
		\STATE $\bs^t=\bnu_{\bs}\cdot\left(\br-\bp\right)$\\		
		\STATE  $\bnu_{\bq}=\textbf{1}./\left(\left|\bH^H\right|^2\bnu_s\right)$\\
		\STATE $\bq=\hat \bx^t+\bnu_{\bq}\cdot\bH^H \bs^t$\\
		\STATE 	$\forall j: \xi_{j,a}=\exp\left({-\nu_{q_j}^{-1}|\alpha_a- q_j|^2}\right)$\\
		\STATE 	$\forall j: \beta_{j,a}={\xi_{j,a}}/{\sum\nolimits_{a=1}^{|\mA|}\xi_{j,a}}$\\
		\STATE 	$\forall j:  \hat x^{t+1}_j=\sum_{a=1}^{|\mA|}\alpha_a \beta_{j,a}  $\\
		\STATE 	$\forall j: \nu^{t+1}_{x_j}=\sum_{a=1}^{|\mA|}\beta_{j,a}|\alpha_a- \hat x_j|^2$\\
		\STATE  $t=t+1$
	\end{algorithmic}
	\textbf{Until terminated}
	\label{AMPv1}
\end{algorithm}

\begin{eqnarray}
\hat x_j&=&\sum_{a=1}^{|\mA|}\alpha_a \beta_{j,a}  \\
\nu_{x_j}&=&\sum_{a=1}^{|\mA|}\beta_{j,a}|\alpha_a- \hat x_j|^2,
\end{eqnarray}
where
\begin{equation}
\beta_{j,a}={\xi_{j,a}}/{\sum\nolimits_{a=1}^{|\mA|}\xi_{j,a}}
\end{equation}
with
\begin{equation}
\xi_{j,a}=\exp\left({-\nu_{q_j}^{-1}|\alpha_a- q_j|^2}\right).
\end{equation}

In Algorithm 1, the results $\hat x_{j}$ and $\nu_{x_j}$ in the last iteration can be used for de-mapping, which can be performed based on the pseudo model
\begin{equation}
\hat x_j
=
x_j + w_j,~ j=1,..., MN,
\label{q_n}
\end{equation}
where $w_{j}$ is a Gaussian noise with mean zero and variance $\nu_{x_j}$.

There are two issues we need to point out. First, in the above algorithm the noise variance $\epsilon^{-1}$ is assumed to be known, which, however, is often unknown in practice. Second, it is noted that the deviation of the channel matrix $\bH$ from an i.i.d. (sub-)Gaussian matrix may lead to performance loss of the AMP algorithm, motivating us to employ a robust variant to AMP, e.g., UAMP. In the next sub-section, we address both issues and develop a UAMP-based detection algorithm, where the estimation of the noise variance is incorporated. In particular, we will show how the UAMP algorithm can take advantage of the structure of the channel matrix and implemented with the 2D FFT algorithm, which is much more efficient than the AMP-based algorithm.


\subsection{UAMP-Based Detector with 2D FFT}

We first examine the property of the channel matrix $\bH$ in \eqref{eq:IdealFracY2}, which can be represented as
\begin{eqnarray}
\boldsymbol{H}=\sum_{i=0}^{P-1}\sum_{c=-N_i}^{N_i}\boldsymbol{I}_N(-[c-k_i]_N)\otimes \Big[\boldsymbol{I}_M(l_i)h_i
\nonumber\\
\times\Big(\frac{1-e^{-j2\pi(-c-\kappa_i)}}{N-Ne^{-j2\pi\frac{-c-\kappa_i}{N}}}\Big) e^{-j2\pi\frac{l_i(k_i+\kappa_i)}{MN}}\Big]
\label{eq:IdealFracH}
\end{eqnarray}
where $\boldsymbol{I}_N(-[q-k_i]_N)$ denotes an $N \times N$ matrix obtained by circularly shifting the rows of the identity matrix by $-[q-k_i]_N$, and $\boldsymbol{I}_M(l_i)$ is obtained similarly.
Without fractional Doppler, i.e., $\kappa_i=0$, the channel matrix $\bH$ is reduced to
\begin{eqnarray}
\boldsymbol{H}=\sum_{i=0}^{P-1}\boldsymbol{I}_N(k_i)\otimes\left[\boldsymbol{I}_M(l_i)h_i e^{-j2\pi\frac{l_ik_i}{MN}}\right].
\label{eq:IdealIntH}
\end{eqnarray}

It is noted that $\boldsymbol{H}$ in both \eqref{eq:IdealFracH} and \eqref{eq:IdealIntH} is a block circulant matrix with circulant blocks (BCCB). As a toy example, suppose an OTFS system with $M=4$, $N=3$, $N_i=1$, and $P=3$, where the channel gain vector is $[h_0,h_1,h_2]$, fractional Doppler shifts are {$[-0.1,0.1,0.2]$, and the indices of delay and Doppler taps are $[0,1,2]$ and $[1,3, 4]$, respectively. Then the channel matrix $\bH$ can be expressed as
\begin{eqnarray}
\bH=
\left(
\begin{matrix}
\bH_0 & \bH_2 & \bH_1 \\
\bH_1 & \bH_0 & \bH_2 \\
\bH_2 & \bH_1 & \bH_0
\end{matrix}
\right)_{MN\times MN}\label{eq:bccb}
\end{eqnarray}
with
\begin{eqnarray}
&&\bH_0=h_0g(1,-0.1)\bI_M(0)+h_1g(0,0.1)\bI_M(1) \nonumber\\
&&\ \ \ \ \ \ \ \ + h_2g(1,0.2)\bI_M(2)\nonumber\\
&&\bH_1=h_0g(0,-0.1)\bI_M(0)+h_1g(-1,0.1)\bI_M(1)\nonumber\\
&&\ \ \ \ \ \ \ \ + h_2g(0,0.2)\bI_M(2)\nonumber\\
&&\bH_2=h_0g(-1,-0.1)\bI_M(0)+h_1g(1,0.1)\bI_M(1) \nonumber\\
&&\ \ \ \ \ \ \ \ + h_2g(-1,0.2)\bI_M(2), \nonumber
\end{eqnarray}
                }
where the function $g(c,\kappa_i)$ is defined as
\begin{eqnarray}
g(c,\kappa_i)=\frac{1}{N}\frac{1-e^{-j2\pi(-c-\kappa_i)}}{1-e^{-j2\pi\frac{-c-\kappa_i}{N}}} e^{-j2\pi\frac{l_i(k_i+\kappa_i)}{MN}}.
\end{eqnarray}


A useful property of the BCCB matrix $\bH$ is that, it can be diagonalized using {a} 2D DFT matrix, i.e.,
\begin{align}
\bH=\bF^H\bLambda\bF \label{eq:IdealDiag}
\end{align}
where $\bF=\bF_N\otimes \bF_M$ {is a unitary matrix} with $\bF_N$ and $\bF_M$ denoting respectively the normalized $N$-point and $M$-point DFT matrices, and matrix $\bLambda$ in \eqref{eq:IdealDiag} is a diagonal matrix, i.e.,
\begin{align}
\bLambda = \text{diag}(\bd) \label{eq:IdealLmd}
\end{align}
with $\bd$ being a length-$MN$ vector. The vector $\bd$  can be computed as
\begin{eqnarray}
\bd=\text{vec}(\text{FFT2}\left(\bC\right)) \label{eq:Diag_FFT},
\end{eqnarray}
where FFT2 ($\cdot$) represents the 2D FFT operation, $\bC=\text{reshape}_M\left(\bH(:,1)\right)$ is an $M \times N$ matrix, and $\bH(:,1)$ with length-$MN$ is the first column of matrix $\bH$.


The above property can be exploited by UAMP, leading to a more efficient detection algorithm while with significantly enhanced performance, compared to the AMP-based detector. Instead of using model \eqref{eq:IdealFracY2} directly, the UAMP algorithm \cite{Guo2015UtAMP} works with the unitary transform of the model. Because the channel matrix $\bH$ admits  \eqref{eq:IdealDiag}, we have the following unitary transform of the OTFS system model \eqref{eq:IdealFracY2}
\begin{eqnarray}
\br=\bLambda\bF\bx + \bomega', \label{eq:VectorY2}
\end{eqnarray}
where $\br\triangleq\bF\by$, $\bomega'\triangleq\bF\bomega$, and the noise $\bomega'$  has the same distribution as $\bomega$ since $\bF$ is an unitary matrix. The precision of the noise is still denoted by $\epsilon$, which needs to be estimated.

\begin{figure}[htbp]
	\centering
	\includegraphics[width=0.85\columnwidth]{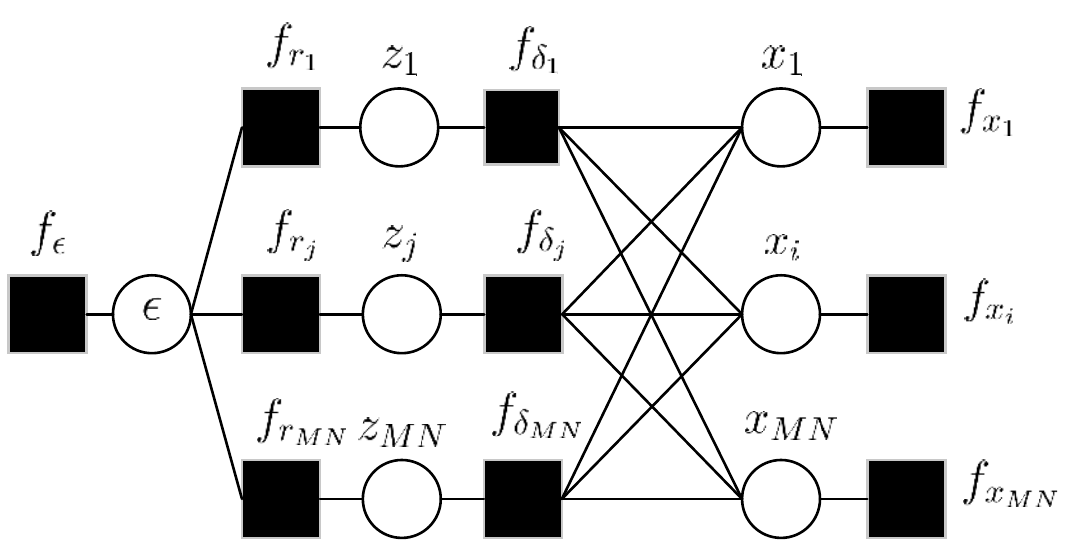}
	\caption{Factor graph representation of \eqref{eq:fact}.}
	\label{fig:fg}
\end{figure}

By defining $\bPhi\triangleq\bLambda\bF$ and an auxiliary vector $\boldsymbol{z}\triangleq \bPhi\bx$,
we can factorize the joint probability density function of the unknown variables $\bx, \bz,$ and  $\epsilon$ given $\br$ as
\begin{eqnarray}
\!\!\!\!\!\!p(\bx, \bz,  \epsilon|\br)\!\!\!\!\!\!\!\!\!\!\!\!\!\ && \propto p(\epsilon)p(\br|\bz,\epsilon)p(\bz|\bx)p(\bx)\nonumber\\
&& =p(\epsilon)\prod\nolimits_{j}p(r_j|z_j,\epsilon)p(z_j|\bx)
\prod\nolimits_{i}p(x_i)\nonumber\\
&& =f_{\epsilon}\prod\nolimits_{j}f_{r_j}(z_j,\epsilon)f_{\delta_j}(z_j,\bx)\prod\nolimits_{i}f_{x_i}(x_i), \label{eq:fact}
\end{eqnarray}
where indexes $i,j\in[1:MN]$. To facilitate the factor graph representation of the factorization in \eqref{eq:fact}, we introduce the notations in Table I, showing the correspondence between the factor labels and the underlying distributions they represent, and the specific functional form assumed by each factor. In Table I, $\bPhi_j$ denotes the $j$th row of matrix $\bPhi$. The factor graph representation for the factorization in \eqref{eq:fact} is depicted in Fig. \ref{fig:fg}, where squares and circles represent function nodes and variable nodes, respectively.

\begin{table}[!hbp]
	\centering
	\caption{Factors, Underlying Distributions and Functional Forms Associated with \eqref{eq:fact}}
	\begin{tabular}{lll}
		\hline
		Factor & Distribution & Functional Form\\
		\hline
		$f_{r_j}$ & $p\left(r_j|z_j,\epsilon\right)$ & $\N\left(z_j;r_j,\epsilon^{-1} \right)$\\
		$f_{\delta_j}$ & $p\left(z_j|\bx\right)$ & $\delta\left(z_j-\bPhi_j \bx\right)$ \\
		$f_{x_i}$ & $p\left(x_i\right)$ & $(1/|\mA|)\sum_{a=1}^{|\mA|} \delta\left(x_i-\alpha_a\right)$ \\
		$f_{\epsilon}$ & $p(\epsilon)$ & $\propto\epsilon^{-1}$ { \cite{Tipping}} \\
		\hline
	\end{tabular}
\end{table}

\begin{algorithm}
	\setstretch{1.25}
	\caption{UAMP Based OTFS Detector (with noise precision estimation)}
	Unitary transform: $\br=\bF\by=\bLambda\bF\bx + \bomega$, with $\boldsymbol{F}=\boldsymbol{F}_N\otimes \bF_M$. Calculated $\bd$ with \eqref{eq:Diag_FFT}, and define vector $\blambda=\bd\cdot\bd^*$. Initialize $\bs^{(-1)}=\boldsymbol{0}$, $\hat\bx^{(0)}=\boldsymbol{0}$, $\hat\epsilon^{(0)}=1$, $\nu_x^{(0)}=1$, and $t=0$.\\
	\textbf{Repeat}
	\begin{algorithmic}[1]
		\STATE $\bnu_{\bp}=\nu_{x}^t\blambda$ \label{eq:UTnp}\\
		\STATE $\bp=\bd\cdot \text{vec}\left(\text{FFT2}\left(\text{reshape}_M
		(\hat\bx^{t})\right)\right)-\bnu_{\bp}\cdot\bs^{t-1}$\label{eq:UTp}\\
		\STATE $\bnu_{\bz}=1./\left(1./\bnu_{\bp}+\hat \epsilon^{t}\bone\right)$\label{eq:UTnz}\\
		\STATE $\hat \bz=\bnu_{\bz}\cdot\left(\bp./\bnu_{\bp}+\hat \epsilon^t\br\right)$\\
		
		\STATE 	$\hat \epsilon^{t+1}={MN}/\left({\left\|\br-\hat \bz\right\|^2_2+\textbf{1}^T\bnu_{\bz}}\right)$\label{eq:epsilon}\\
		
		\STATE 	$\bnu_{\bs}=1./\left(\bnu_{\bp}+1/\epsilon^{t+1}\boldsymbol{1}\right)$\\
		\STATE 	$\bs^t=\nu_{\bs}\cdot\left(\br-\hat \bp\right)$\\
		
		\STATE 	$\nu_{q}=\blambda^T\bnu_s/(MN)$\\
		\STATE 	$\bq=\hat \bx^{(t)}+\nu_{q} \text{vec}\left(\text{IFFT2}\left(\text{reshape}_M({\bd}\cdot\bs^t)\right)\right)$ \label{eq:UTq}\\
		
		\STATE 	$\forall j: \xi_{j,a}=\exp\left({-\nu_{q}^{-1}|\alpha_a- q_j|^2}\right)$ \label{eq:UTbeta}\\
		\STATE 	$\forall j: \beta_{j,a}={\xi_{j,a}}/{\sum\nolimits_{a=1}^{|\mA|}\xi_{j,a}}$\\
		
		\STATE 	$\forall j: \hat  x^{t+1}_j=\sum\nolimits_{a=1}^{|\mA|}\alpha_a \beta_{j,a}$ \label{eq:UTx}\\
		\STATE 	$\forall j: \nu_{x_j}^{t+1}=\sum\nolimits_{a=1}^{|\mA|}\beta_{j,a}|\alpha_a- \hat x^{t+1}_j|^2$ \label{eq:UTnx}\\
		\STATE 	$\nu_x^{t+1}=\frac{1}{MN}\sum_{j=1}^{MN}\nu_{x_j}^{t+1}$\\
		\STATE $t=t+1$
	\end{algorithmic}
	\textbf{Until terminated}
	\label{UTAMPv1}
\end{algorithm}

Following the UAMP {algorithm}, we can derive a UAMP-based iterative detector, which is summarized in Algorithm \ref{UTAMPv1}. {As the noise precision $\epsilon$ is unknown}, its estimation needs to be included in the UAMP-based detector. According to the derivation of (U)AMP using the loopy belief propagation, UAMP provides the message from variable node $z_j$ to function node $f_{r_j}$, which is Gaussian and denoted by  $m_{{z_j}\rightarrow f_{r_j}} (z_j) = \mathcal{N} (z_j |  p_j, \nu_{p_j})$. Here, the mean $p_j$ and the variance $\nu_{p_j}$ are given in Lines 1 and 2 of Algorithm 2 in a vector form. Based on the mean field rule \cite{combine} at the function node $f_{r_j}$, we can compute the message passed from function node $f_{r_j}$ to variable node $\epsilon$, i.e.,
\begin{equation}
\begin{aligned}
m_{f_{r_j}\rightarrow \epsilon} (\epsilon) & \propto \exp \left\{  \left\langle {\log  f_{r_j} ( {{r_j| z_j, \epsilon}})} \right\rangle_{b\left( {z_j} \right) } \right\}\\
&  \propto \epsilon \exp \left\{   { -\epsilon } { ( |  r_j - \hat{z}_j   |^2 + v_{z_j}  ) } \right\},
\end{aligned}
\end{equation}
where $b(z_j)$ is the belief of $z_j$. It turns out that $b(z_j)$ is also Gaussian with variance and mean given by
\begin{eqnarray}
\nu_{z_j}&=&1/\left(1/\nu_{p_j}+\hat \epsilon\right)
\end{eqnarray}
and
\begin{eqnarray}
\hat z&=&\nu_{z_j}\left(p_j/\nu_{p_j}+\hat \epsilon r_j\right)
\end{eqnarray}
respectively, where $\hat \epsilon$ in the estimate of $\epsilon$ in last iteration. They can be expressed in a vector form shown in Line 3 and Line 4 of Algorithm \ref{UTAMPv1}.
The estimation of $\epsilon$ can be performed based on the belief $b(\epsilon)$ at the variable node $\epsilon$ shown in Fig. \ref{fig:fg}, i.e.,
\begin{equation}
\begin{aligned}
b(\epsilon)& \propto f_{\epsilon} (\epsilon)  \prod_{j=1}^{MN} m_{f_{r_j}\rightarrow \epsilon} (\epsilon). \\
\end{aligned}
\end{equation}
Then the estimate of $\epsilon$ can be expressed as
\begin{equation}
\hat{\epsilon}  = \int_{0}^{\infty} \epsilon b(\epsilon)d\epsilon= {MN}/{\sum_{j=1}^{MN} { \left( | r_j - \hat{z}_j  |^2 + \nu_{z_j}  \right) }},\label{n1}
\end{equation}
which can be rewritten in a vector form shown in Line 5 of Algorithm \ref{UTAMPv1}. By using the mean field rule at the function node $f_{r_j}$ again, we have the message passed from the function node $f_{r_j}$ to the variable node $z_j$, i.e.,
\begin{equation}
\begin{aligned}
m_{f_{r_j}\rightarrow z_j} (z_j)
& \propto \exp \left\{  \left\langle {\log  f_{r_j} ( {{r_j| z_j, \hat\epsilon}}  )} \right\rangle_{b\left( {\epsilon} \right) } \right\}\\
&\propto \mathcal{N} (h_j |  r_j, \hat{\epsilon}^{-1}).
\label{M_frm_to_hm}
\end{aligned}
\end{equation}
Then the UAMP algorithm with known noise can be used as if the true noise precision is $\hat{\epsilon}$, leading to Lines 6-15 and Lines 1 and 2 of  Algorithm \ref{UTAMPv1}. The derivations of Lines 10-13 are the same as those for the AMP-based detector. There is an extra operation in Line 14, which averages the variances of $\{x_j\}$. It is note that the special form of the unitary matrix $\bF$ leads to the implementation with the 2D FFT algorithm in Line 2 and Line 9 of Algorithm \ref{UTAMPv1}.

\subsection{Computational Complexity Analysis}
We first look at the complexity of the AMP based-detector. From Algorithm 1, we can find that the complexity of AMP-based detector is dominated by the matrix-vector products in Lines 1, 2, 5 and 6, e.g., $\bH\hat\bx^t$ and $\left|\bH\right|^2\bnu_{\bx}^t$. Note that, each row of the matrix $\bH$ has $S=\sum_{i=1}^P N_i$ non-zero elements \cite{Raviteja2018},  so the complexity is $\mO\left(MNS\right) + \mO\left(MN|\mA|\right)$ per OTFS block per iteration.

Comparing Algorithm 2 for the UAMP-based detector with Algorithm 1, we can find that the UAMP-based detector does not require any matrix-vector products, and all lines of the algorithm involve only element-wise vector operations or scalar operations, except Line 2 and Line 9. It is noted that Line 2 and Line 9 can be implemented with the FFT algorithm. So the the complexity of Algorithm 2 is $\mO\left(MN\log(MN)\right) +\mO\left(MN|\mA|\right)$ per OTFS block per iteration, which is independent of $S$.

In comparison with the (U)AMP-based detectors, the detectors proposed in \cite{Raviteja2018} and \cite{Yuan2019} have a complexity of  $\mO(MNS|\mA|)$ per OTFS block per iteration, which can be much higher than that of the (U)AMP-based detectors in the case of rich scattering environments and when fractional Doppler shifts have to be considered (leading to a large $S$). {The MRC detector in \cite{rakea, rake} has a complexity of $\mO(MN(L+log(N)))$ per block per iteration ($L$ is the number of unique delay taps), which can be smaller than that of the UAMP-based detector when $L$ is small.}  We will show in Section VI that the (U)AMP-based detectors can deliver much better performance {than other detectors} especially when $P$ is relatively large.

\section{OTFS Detection with Rectangular Waveform}

{The rectangular waveform has been used in the literature as a more practical waveform. When a length-$T$ rectangular waveform is used as the transmitter and receiver filters, as in \cite{Ahmad2017}, \cite{OTFSCP, channelofdm}, we append a cyclic prefix (CP) to each length-$M$ sub-block of the signal $\bs$ in the time domain before transmitting over the time-varying channel.

After removing CPs at the receiver side, the received signal in the time domain can be expressed as \cite{Ahmad2017,OTFSCP,channelofdm}
\begin{eqnarray}
\boldsymbol{u}
=\bH_T \bs+\bomega,
\end{eqnarray}
where $\bomega$ is an additive Gaussian white noise vector, and $\bH_T$ is an $NM\times NM$ block diagonal matrix
\begin{eqnarray}
\bH_{T}=
\left(
 \begin{matrix}
   \bH_1 & 0 & \ldots& 0 \\
   0 & \bH_2 & \ldots &0 \\
   \vdots & \vdots & \ddots &\vdots \\
   0 & 0 &\ldots & \bH_N
  \end{matrix}
  \right)_{NM\times NM} \nonumber 
\end{eqnarray}
where $\bH_n\in\mathbb{C}^{M\times M}$ is the channel matrix corresponding to the $n$-th sub-block of the transmitted signal. Then the received signal in the DD domain is obtained by performing block-wise DFTs, followed by the SFFT operation, i.e.,
\begin{align}
  \by = (\bF_N\otimes \bF_M^H)(\bI \otimes  \bF_M)\boldsymbol{u}. \label{eq:rectddy}
\end{align}
Finally, the channel input-output relationship in the DD domain can be expressed as
\begin{eqnarray}
\by&=&(\bF_N\otimes \bI_M)\bH_T(\bF_N^H\otimes \bI_M)\bx+\bomega\nonumber\\
&=&\bH\bx+\bomega, \label{eq:RectY}
\end{eqnarray}
where $\bH$ represents the effective channel matrix in the DD domain.
}

Based on model \eqref{eq:RectY}, we can directly apply Algorithm 1 to detect $\bx$, leading to an AMP-based OTFS detector. However, Algorithm 2 cannot be directly applied.  Next, we design a detector based on the UAMP algorithm.

Since the channel matrix $\bH_T$ is a block diagonal matrix, we can perform SVD block-by-block, i.e.,
\begin{eqnarray}
\bH&=&
\left(
 \begin{matrix}
   \bU_1\bLambda_1\bV_1 & 0 & \ldots& 0 \\
   0 & \bU_2\bLambda_2\bV_2 & \ldots &0 \\
   \vdots & \vdots & \ddots &\vdots \\
   0 & 0 &\ldots & \bU_N\bLambda_N\bV_N
  \end{matrix}
  \right)\nonumber\\
  &=&\bU\bLambda\bV
  \label{eq:RectH}
\end{eqnarray}
where
\begin{eqnarray}
\bU&=&
\left(
\begin{matrix}
\bU_1 &  \ldots& 0 \\
\vdots & \ddots &\vdots \\
0 & \ldots & \bU_N
\end{matrix}
\right)
\end{eqnarray}
\begin{eqnarray}
\bLambda&=&
\left(
 \begin{matrix}
   \bLambda_1 &  \ldots& 0 \\
   \vdots & \ddots &\vdots \\
   0 & \ldots & \bLambda_N
  \end{matrix}
  \right) \\
  \bV&=&
\left(
 \begin{matrix}
   \bV_1 &  \ldots& 0 \\
   \vdots & \ddots &\vdots \\
   0 & \ldots & \bV_N
  \end{matrix}
  \right),
\end{eqnarray}
and each $\bLambda_n$ is an $M \times M$  diagonal matrix (so does $\bLambda$ itself).
By applying the unitary transformation with the unitary matrix { $\bU^H(\bF_N^H\otimes \bI_M)$} to model \eqref{eq:RectY}, we have
\begin{eqnarray}
\br=\bLambda\bPhi\bx+\bar\bomega, \label{eq:RectY2}
\end{eqnarray}
where $\br=\bU^H(\bF_N^H\otimes \bI_M)  \by\in\mathbb{C}^{MN\times 1}$, $\bPhi=\bV(\bF_N^H\otimes \bI_M)$ and
$\bar\bomega=\bU^H(\bF_N^H\otimes \bI_M)\bomega$ is still a zero-mean Gaussian noise vector with the same covariance matrix as $\bomega$ because $\bU^H(\bF_N^H\otimes \bI_M)$ is a unitary matrix. So, UAMP can be applied to model \eqref{eq:RectY2}, leading to an algorithm similar to Algorithm 2 with a few equations modified, as detailed in the following:
\begin{itemize}
\item[a.] Vector $\bd$ consists of the diagonal elements of matrix $\bLambda$.
\item[b.] The computation of $\bp$ in Line \ref{eq:UTp} of Algorithm 2 is replaced by
\begin{eqnarray}
\bp&=&\bLambda\bV(\bF_N^H\otimes \bI_M)\hat\bx-\bnu_p\cdot\bs^{t-1}\nonumber\\
&=&\bd\cdot\bV\text{vec}(\hat\bX\bF_N^H)-\bnu_p\cdot\bs^{t-1},
\label{eq:RectHatp}
\end{eqnarray}
where $\hat\bX\triangleq \text{reshape}_M(\hat\bx)$, and $\hat\bX\bF_N^H$ can be computed using the FFT algorithm. 
\item[c.] The computation of $\bq$ in Line \ref{eq:UTq} of Algorithm 2 is replaced by
\begin{eqnarray}
\bq&=&\hat\bx^t+\nu_q (\bLambda\bPhi)^H\bs\nonumber\\
&=&\hat\bx^t+\nu_q (\bF_N\otimes \bI_M)\bV^H\bLambda^H\bs\nonumber\\
&=&\hat\bx^t+\nu_q\text{vec}(\bS\bF_N),\label{eq:RectHatq}
\end{eqnarray}
where $\bS\triangleq \text{reshape}_M(\bV^H(\bd^*\cdot\bs))$, and $\bS\bF_N$ can be computed using the FFT algorithm.
\end{itemize}
The UAMP-based detector with rectangular waveform only involves element-wise product (division) except \eqref{eq:RectHatp} and \eqref{eq:RectHatq}. Note that $\bV$ is a block diagonal matrix, so the computational complexity of the detector is $\mO\left(M^2N\right)+\mO(MN|\mA|)$ per OTFS block. 
{which is independent of $S$. Compared to the MRC detector in \cite{rakea, rake}, the UAMP-based detector has a higher complexity, but as shown in Section VI, it can achieve significantly better performance, especially in the case of high order modulations.}

\begin{figure*}[htbp]
	\centering
	\includegraphics[width=1.5\columnwidth]{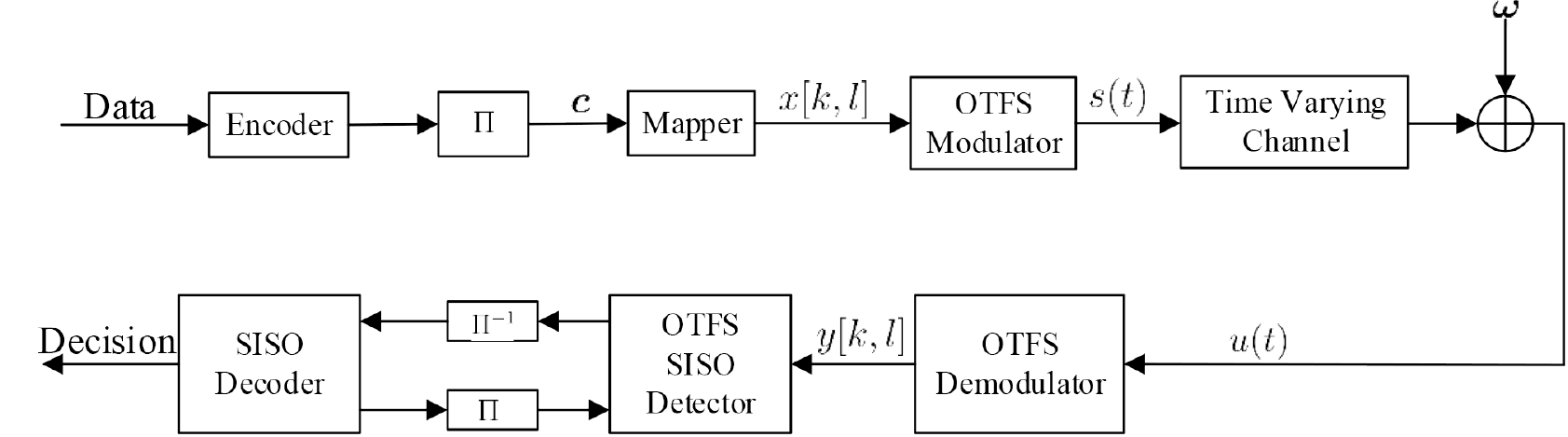}
	\caption{Iterative joint detection and decoding in a coded OTFS system, where II and II$^{-1}$ represent interleaver and deinterleaver, respectively.}
	\label{fig:iterative}
\end{figure*}

\section{{Extension to Coded System with Turbo Receiver and BER Performance Prediction with State Evolution}}

In this section, {we consider a coded OFTS system and investigate the design of a turbo receiver for joint detection and decoding.}

The turbo system is shown in Fig. \ref{fig:iterative}, where the information bits are firstly encoded and then interleaved before mapping. Each symbol $x_j \in\mA=\{\alpha_1, ..., \alpha_{|\mA|}\}$ in the DD domain is mapped from a sub-sequence of the coded bit sequence, which is denoted by $\bc_j=[c_j^1, ..., c_j^{log|\mA|}]$. Each $\alpha_a$ corresponds to a length-$\mathrm{log}|\mA|$ binary sequence denoted by $\{\alpha_a^1, ...,\alpha_a^{log|\mA|}\}$. The turbo receiver consists of a soft-in-soft-out (SISO) detector and a SISO decoder, which exchange information in an iterative manner. Based on the log-likelihood ratios (LLRs) provided the SISO decoder and the output of the OTFS demodulator as shown in Fig. \ref{fig:iterative}, the task of the SISO detector is to compute the extrinsic LLR for each coded bit, i.e.,
{
\begin{equation}
L^e(c_j^q)=\ln\frac{P(c_j^q=0|\br)}{P(c_j^q=1|\br)}-L^a(c_j^q),
\end{equation}
where $L^a(c_j^q)$ is the ouput extrinsic LLR of the decoder in last iteration. The extrinsic LLR  $L^e(c_j^q)$ is passed to the decoder. The derivation for $L^e(c_j^q)$ in terms of extrinsic mean and variance can be find in \cite{GuoAConcise}, and {$L^e(c_j^q)$ can be expressed as}
	\begin{equation}
	L^e(c_j^q)=\ln\frac{\sum_{\alpha_a\in \mA_q^0} \exp\big(-\frac{|\alpha_a-m_j^e|^2}{v_j^e}\big)\prod_{q'\neq q}P(c_j^{q'} = \alpha_a^{q'})}{\sum_{\alpha_a\in \mA_q^1} \exp\big(-\frac{|\alpha_a-m_j^e|^2}{v_j^e}\big)\prod_{q'\neq q}P(c_j^{q'} = \alpha_a^{q'})}, \label{eq:new}
	\end{equation}
where $m_j^e$ and $v_j^e$ are the extrinsic mean and variance of $x_j$, and
$\mA_q^0$ and $\mA_q^0$ represent the subsets of all $\alpha_a$ corresponding to $c_j^q=0$  and $c_j^q=1$, respectively. The extrinsic variance and mean are defined as \cite{GuoAConcise}
	\begin{eqnarray}
	v_j^e &=& (1/v_j^p - 1/v_j)^{-1} \\
	m_j^e &=& v_j^e(m_j^p/v_j^p - m_j/v_j)
	\end{eqnarray}
where $m_j$ and $v_j$ are the \emph{a priori} mean and variance of $x_j$ calculated based on the output LLRs of the SISO decoder \cite{mmseequa}, \cite{Guograph}, \cite{turbobook} and $m_j^p$ and $v_j^p$ are the \emph{a posteriori} mean and variance of $x_j$.} By examining the derivation of the (U)AMP algorithm, we can find that $\bq$ and $\bnu_q$ consists of the extrinsic means and extrinsic variances of the symbols in vector $\bx$ as they are the messages passed from the observation side and do not contain the immediate \emph{a priori} information about $\bx$. Therefore, we have
\begin{eqnarray}
m_j^e= q_j, ~v_j^e= \nu_{q_j}
\end{eqnarray}
in Algorithm 1 and
\begin{eqnarray}
m_j^e= q_j, ~v_j^e= \nu_{q}
\end{eqnarray}
in Algorithm 2.
Then \eqref{eq:new} can be readily used to compute the extrinsic LLRs of the coded bits.

It is noted that, with the LLRs provided by the SISO decoder, we can compute the probability $p(x_j=\alpha_a)$ for each $x_j$, which is no longer $1/|\mA|$ in Algorithm 1 and Algorithm 2. Therefore, $\xi_{j,a}$ in Line 7 of Algorithm 1 is amended to
\begin{eqnarray}
\xi_{j,a}=p(x_j=\alpha_a)\exp\Big({-\nu_{q_j}^{-1}|\alpha_a - q_j|^2}\Big),
\end{eqnarray}
and Line \ref{eq:UTbeta} of Algorithm 2 is changed to
\begin{eqnarray}
\xi_{j,a}=p(x_j=\alpha_a)\exp\Big({-\nu_{q}^{-1}|\alpha_a - q_j|^2}\Big).
\end{eqnarray}

In addition, we note that the iteration of the (U)AMP-based detector can be incorporated into the iteration between the SISO decoder and detector, leading to a single loop iteration (i.e., inner iterations are not required).



{Next}, we investigate how to predict the performance of the (U)AMP-based OTFS detector based on the state evolution (SE). As (U)AMP decouples the estimation of vector $\bx$, so that, in the $t$th iteration, we have the following pseudo observation model
\begin{equation}
q^t_j=x_j+\varpi^t_j, \label{awgn1}
\end{equation}
where $j=1, ..., J$ with $J=MN$,
and $\{\varpi^t_j\}$ denote an additive white Gaussian noise (AWGN) with mean 0 and variance $\tau^t$. In UAMP, the variance of the AWGN $\tau^t$ is given as
\begin{eqnarray}
\tau^t &=& \frac{J}{\bm{1}^T \big(\bm{\lambda}./(v_x^{t}\bm{\lambda}+\epsilon^{-1}\bm{1})\big)} \label{UTAMPSE}
\end{eqnarray}
where $v_x^t$ is the (average) variance of $\{x_j\}$ in the $t$th iteration.

Based on the above, the coded OTFS system can be regarded as a pseudo coded system with a simple AWGN channel give in \eqref{awgn1}, where $\{x_j\}$ are mapped from a coded bit sequence. In the pseudo coded system, $\{q^t_j\}$ are the received signal, and the noise variance is $\tau^t$, i.e., the SNR is $1/\tau^t$, where we assume that the power of the signal is 1. Then the signal is demapped with \eqref{eq:new} and the LLRs are input to the SISO decoder. Based on the output of the decoder, the variance $v_x^{t+1}$ can be obtained as shown in Line 10 of Algorithm 1 and Line 14 of Algorithm 2. Clearly, $v_x^{t+1}$ depends on the SNR or $\tau^t$ of the pseudo AWGN channel, i.e.,
\begin{equation}
v_x^{t+1}=g(\tau^t),
\end{equation}
where $g(\cdot)$ is a function.
It is noted that the function normally does not have an analytical form. However, it can be established (in the form of a table) through simulation as in \cite{snrevolution}, i.e., simulate the code (with the mapper used) in AWGN channels with different SNRs. At the same time the bit error rate (BER) can also be obtained. Hence, we can establish the 'function'
\begin{equation}
(\mathrm{BER}, v_x)=g(\tau).
\end{equation}
It is noted that the 'function' does not depend on the OTFS channel, and it is only related to the code, the symbol mapper used and the SNR of the AWGN channel.

Therefore the BER performance of the coded OTFS system with iterations can be predicated with the following simple iterative process with initialization $t=0$ and  $v_x^{t}=1$.
\begin{eqnarray}
\begin{aligned}
&\!\!\!\!\!\!\!\!\!\!\!\!\!\!\!\mathrm{Repeat} \nonumber \\
&\tau^t = \frac{J}{\bm{1}^T \big(\bm{\lambda}./(v_x^{t}\bm{\lambda}+\epsilon^{-1}\bm{1})\big)} \nonumber \\
&(\mathrm{BER}^{t+1}, v^{t+1}_x)=g(\tau^t) \nonumber \\
&t=t+1 \nonumber \\
&\!\!\!\!\!\!\!\!\!\!\!\!!\!\!\mathrm{Until~terminated} \nonumber
\end{aligned}
\end{eqnarray}
As we will see in Section VI that the BER performance of the coded OTFS system with the UAMP-based detector can be predicted fairly well. {Here we note that this is an empirical finding as rigorous SE of AMP algorithms for a general system transform matrix is still an open problem.}  In contrast, for the system with AMP-based detector, there is a huge difference between the simulated performance and predicted performance. This is because the channel matrix deviates from the i.i.d. Gaussian matrix, and as the AMP SE is no longer valid.

\section{Simulation Results}
In this section, {we evaluate the performance of the UAMP-based detectors and compare them with the AMP-based detectors and the state-of-the-art low complexity detectors in \cite{Raviteja2018}, \cite{rake}}  and \cite{Yuan2019}, {which are named MP, MRC and VB, respectively.
}
We set $M = 256$ and $N=32$, i.e., there are $32$ time slots and $256$ subcarriers in the TF domain. {Both quadrature phase shift keying (QPSK) modulation and 16-quadrature amplitude modulation (QAM) are considered}. The carrier frequency is 3 GHz, and the subcarrier spacing is 2 kHz. The speed of the mobile user is set to be $v = 135 km/h$, leading to a maximum Doppler frequency shift index $k_{max}=6$. We assume that the maximum delay index is {$l_{max}$ = 14}. The Doppler index of the $i$th path is uniformly drawn from the set $[-k_{max}, k_{max}]$ and the delay index belongs to $[1, l_{max}]$ excluding the first path ($\l_1=0$). We assume that the fractional Doppler $\kappa_i$ is uniformly distributed within $[-1/2, 1/2]$, and {the channel coefficients $\{h_i\}$ are independently drawn from a complex Gaussian distribution with mean $0$ and variance $\eta^{l_{i}}$, where the normalized power delay profile $\eta^i={\text{exp}(-\alpha l_i)}/{\sum_i \text{exp}(-\alpha l_i)}$ with $\alpha$ being $0$ or $0.1$ \cite{Yuan2019}, \cite{Guodouble}. }
The maximum number of iterations is set to be 15 for the iterative algorithms. We note that, {all the detectors except the MRC detector require the precision (or variance) of the noise}. {The UAMP-based detectors perform noise precision estimation, while he other detectors (except the MRC detector) including the AMP-based detectors assume that the noise precision is perfectly known. We examine the performance of the detectors in a variety of scenarios including the bi-orthogonal waveform with integer or the fractional Doppler shifts, rectangular waveform with fractional Doppler shifts, and QPSK or 16-QAM employed for modulation. In addition, both uncoded and coded systems are considered.}


\begin{figure}[h]
	\centering
	\includegraphics[width=1\columnwidth]{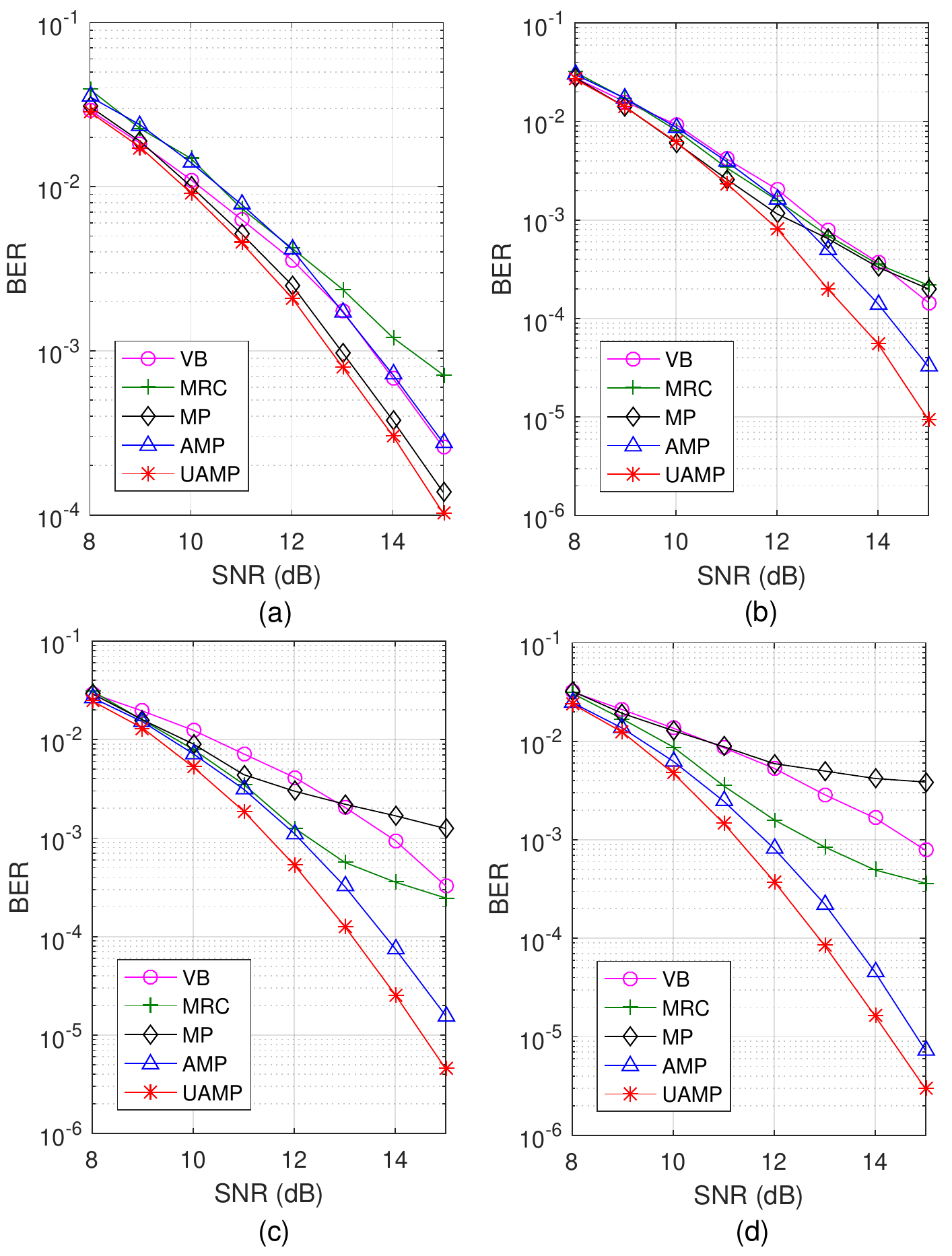}
	\caption{{BER performance of various detectors with bi-orthogonal waveform and integer Doppler shifts (a)$P=6$, (b)$P=10$, (c) $P=12$, and (d) $P=14$.}}
	\label{fig:IdeaIntBerVsSnr}
\end{figure}

\begin{figure}[h]
	\centering
	\includegraphics[width=0.8\columnwidth]{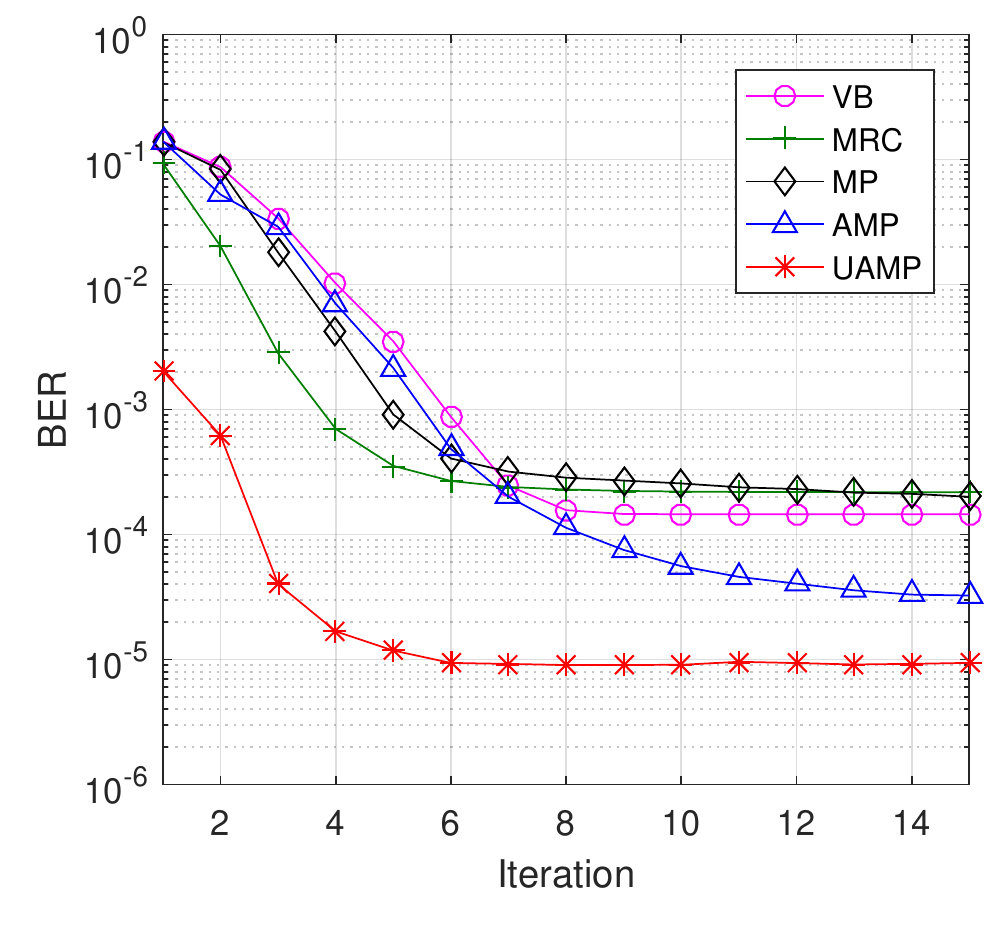}
	\caption{{BER performance versus iteration number (bi-orthogonal waveform and integer Doppler shifts), where $P=10$ and SNR = 15dB.}}
	\label{fig:IdeaIntBerVsIter}
\end{figure}

{We start with the performance comparisons of various detectors in the case of the bi-orthogonal waveform. First we assume that} no fractional Doppler shifts exist, i.e., $\kappa_i=0$ and $S=P$, and the results are shown in Fig. \ref{fig:IdeaIntBerVsSnr}.
{{The BER performance of various detectors versus different values of $P$ is shown in  Fig. \ref{fig:IdeaIntBerVsSnr}, where $\alpha=0$, and QPSK is used.} By comparing the results {in this figure}
we can see that, the MP-based detector performs well when $P=6$, but with the increase of $P$, its performance degrades. {Note that the MP algorithm in \cite{Raviteja2018} is an approximation to loopy belief propagation due to a Gaussian approximation used to reduce complexity. The increase of $P$ significantly affects the performance of the MP-based detector, as a denser channel matrix leads to the presence of short loops with a higher probability. The VB-based detector has the similar trend. The MRC detector performs similarly as the MP-based and VB-based detectors when $P=10$ and delivers better performance than the MP-based and VB-based detectors when $P=12$ and $P=14$.} The AMP and UAMP-based detectors perform well, and they enjoy the diversity gain and achieve better performance with the increase of $P$. In all cases, the UAMP-based detector delivers the best performance.
}
To illustrate the convergence behaviors, we plot the BER performance of various detectors versus the number of iterations in Fig. \ref{fig:IdeaIntBerVsIter} with $P=10$ and SNR $=15$dB. It can be seen that {the UAMP-based detector and the MRC detector {converge faster than} other detectors.}

	\begin{figure}[h]
		\centering
		\includegraphics[width=1\columnwidth]{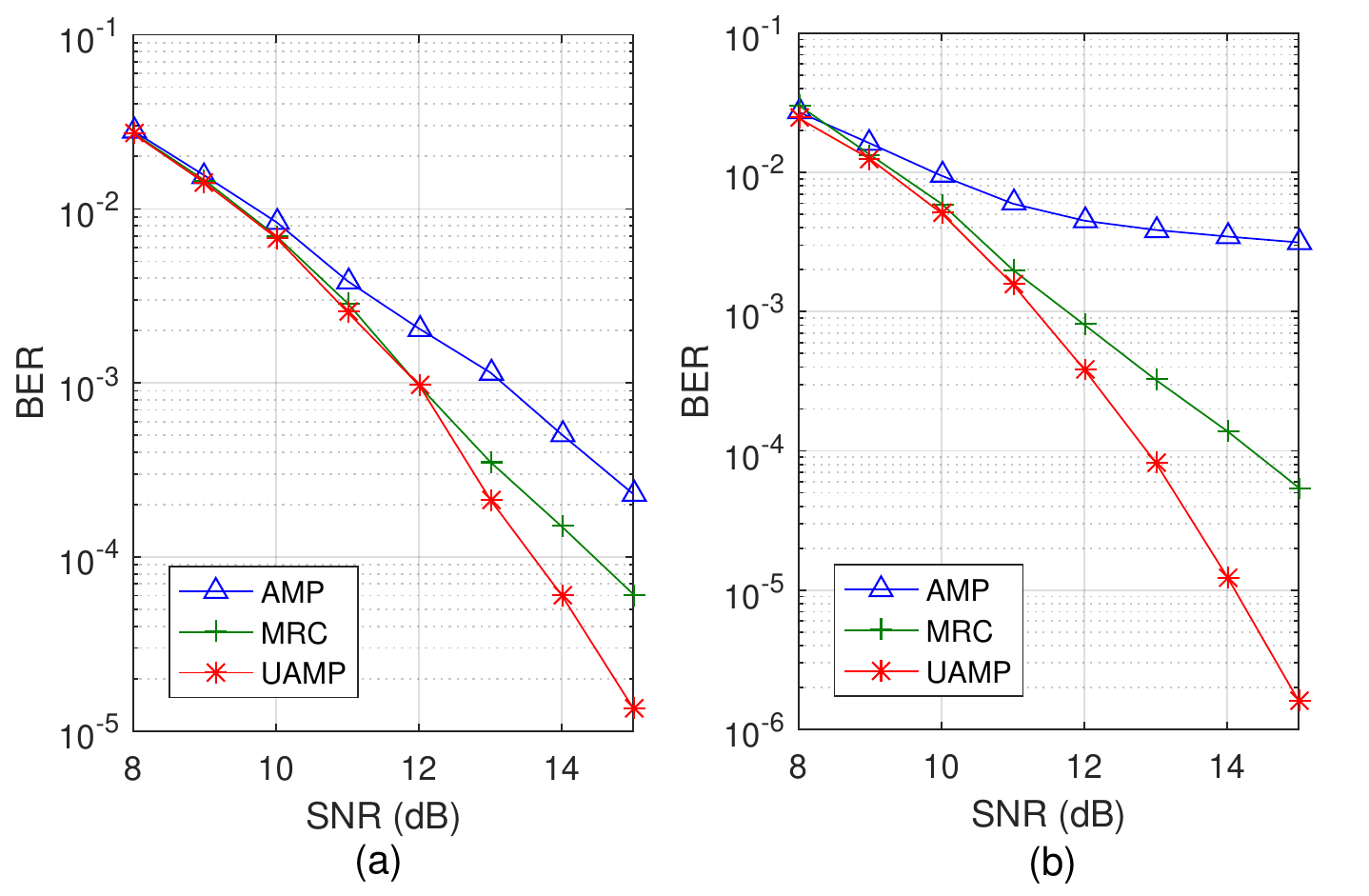}
		\caption{{BER performance of detectors with bi-orthogonal waveform and fractional Doppler shifts (a) $P=10$, and (b) $P=14$.}}
		\label{fig:IdeaFracBerVsSnr}
	\end{figure}
	
	\begin{figure}[h]
		\centering
		\includegraphics[width=1\columnwidth]{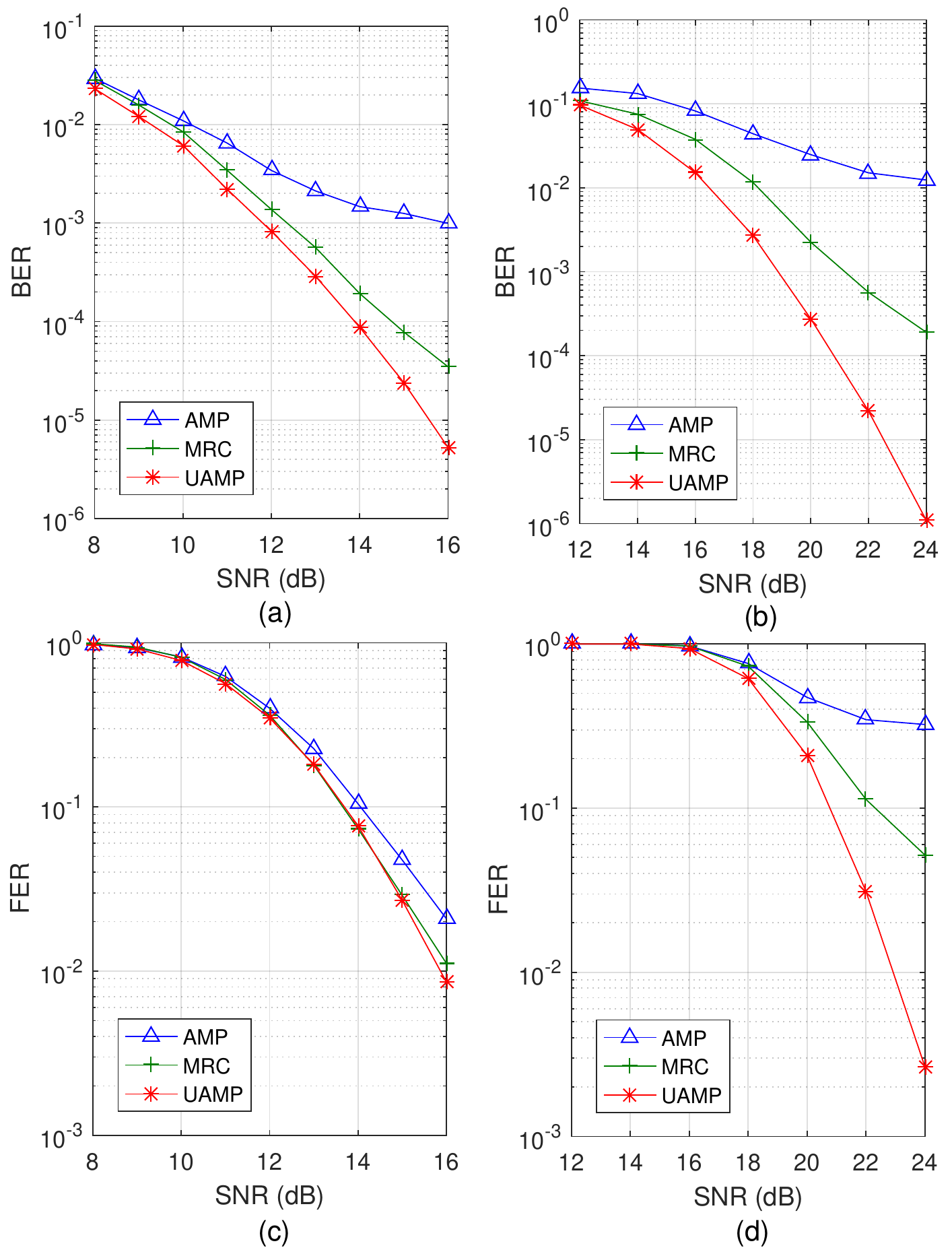}
		\caption{{BER and FER performance of detectors with the rectangular waveform and fractional Doppler shifts (a) QPSK (BER), (b) 16-QAM (BER), (c) QPSK (FER), and (d) 16-QAM (FER).}}
		\label{fig:RectBerVsSnr}
	\end{figure}


	\begin{figure}[h]
		\centering
		\includegraphics[width=1\columnwidth]{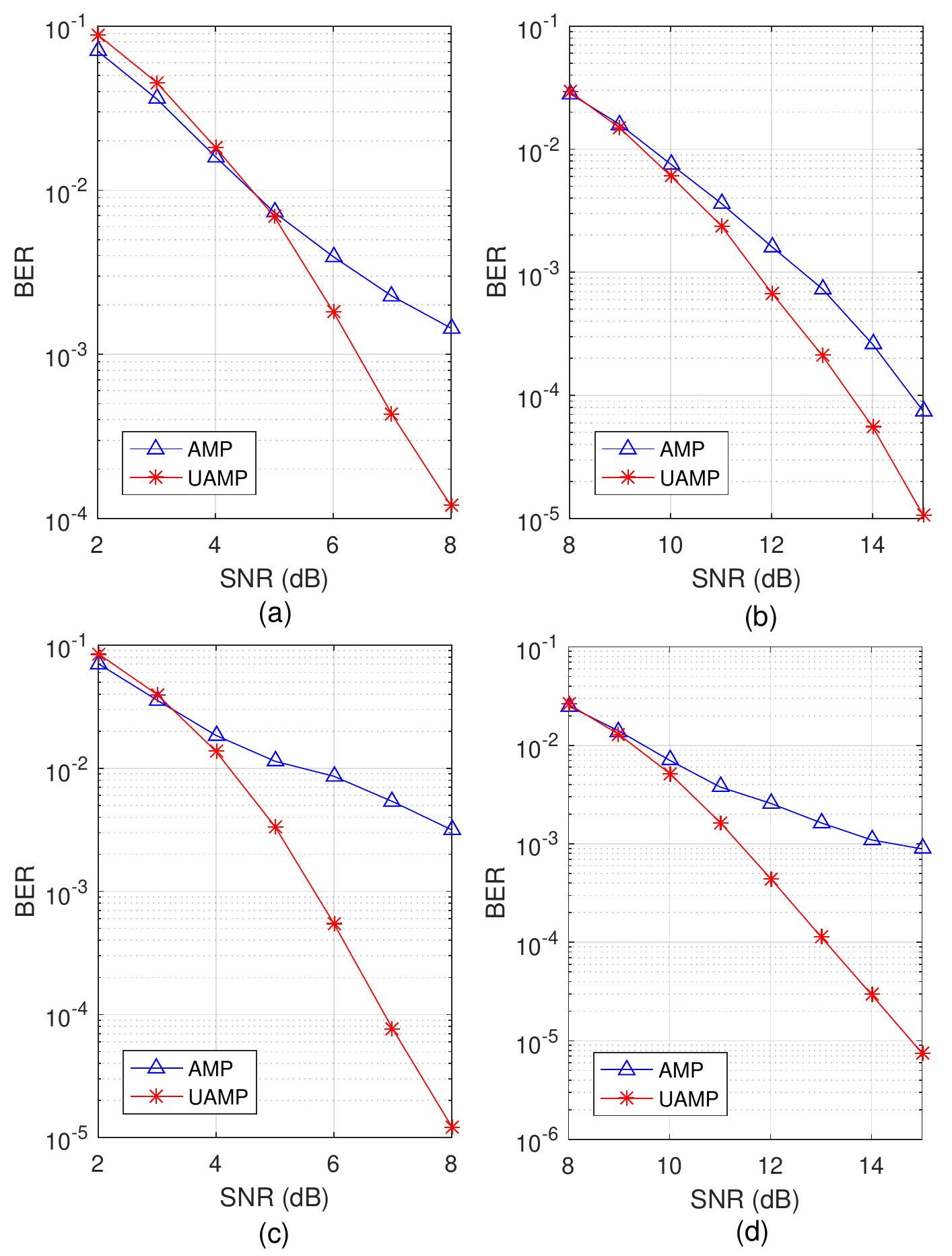}
		\caption {{BER performance of the coded and uncoded systems with the (U)AMP-based detectors with rectangular waveform and fractional Doppler shifts (a) $P=10$ (coded), (b) $P=10$ (uncoded), (c) $P=14$ (coded), (d) $P=14$ (uncoded).}}
		\label{fig:RectBerVsSnrCoded}
	\end{figure}

With the assumption of factional Doppler shifts, we compare the BER performance of the AMP-based detector, the UAMP-based detector {and the MRC detector} versus SNR in Fig. \ref{fig:IdeaFracBerVsSnr}, {where $\alpha=0$ and QPSK is used.} As the channel matrix is not i.i.d. (sub-) Gaussian, UAMP is more robust than AMP, which leads to a significantly better performance for the UAMP-based detector, as demonstrated in Fig. \ref{fig:IdeaFracBerVsSnr}. When $P=14$, the performance of the AMP-based detector gets worse, and there is a huge performance gap between the AMP and the UAMP-based detectors. {The MRC detector delivers much better performance than the AMP-based detector, and the UAMP-based detector
outperforms the MRC detector significantly at relatively high SNRs. }




Next, we consider the rectangular waveform and {compare the performance of the (U)AMP based detectors and the MRC detector versus SNR. We assume fractional Doppler shifts and the number of paths $P=9$, where $\alpha=0.1$ is used for the power delay profile \cite{Yuan2019, Guodouble}. The BER and frame error rate (FER) performance are shown in Fig. \ref{fig:RectBerVsSnr}. It can be seen that, in terms of BER, the MRC detector performs better than the AMP-based detector and the UAMP-based detector still delivers significantly better performance than other detectors, especially in the case of 16-QAM. In terms of FER, the UAMP-based detector and the MRC detector have similar performance in the case of QPSK, and the UAMP-based detector performs considerably better than the MRC detector in the case of 16-QAM. As the MRC detector has lower complexity than the UAMP-based detector, the UAMP-based detector and the MRC detector are two useful options to achieve a good trade-off between performance and complexity.
}

\begin{figure}[h]
	\centering
	\includegraphics[width=0.8\columnwidth]{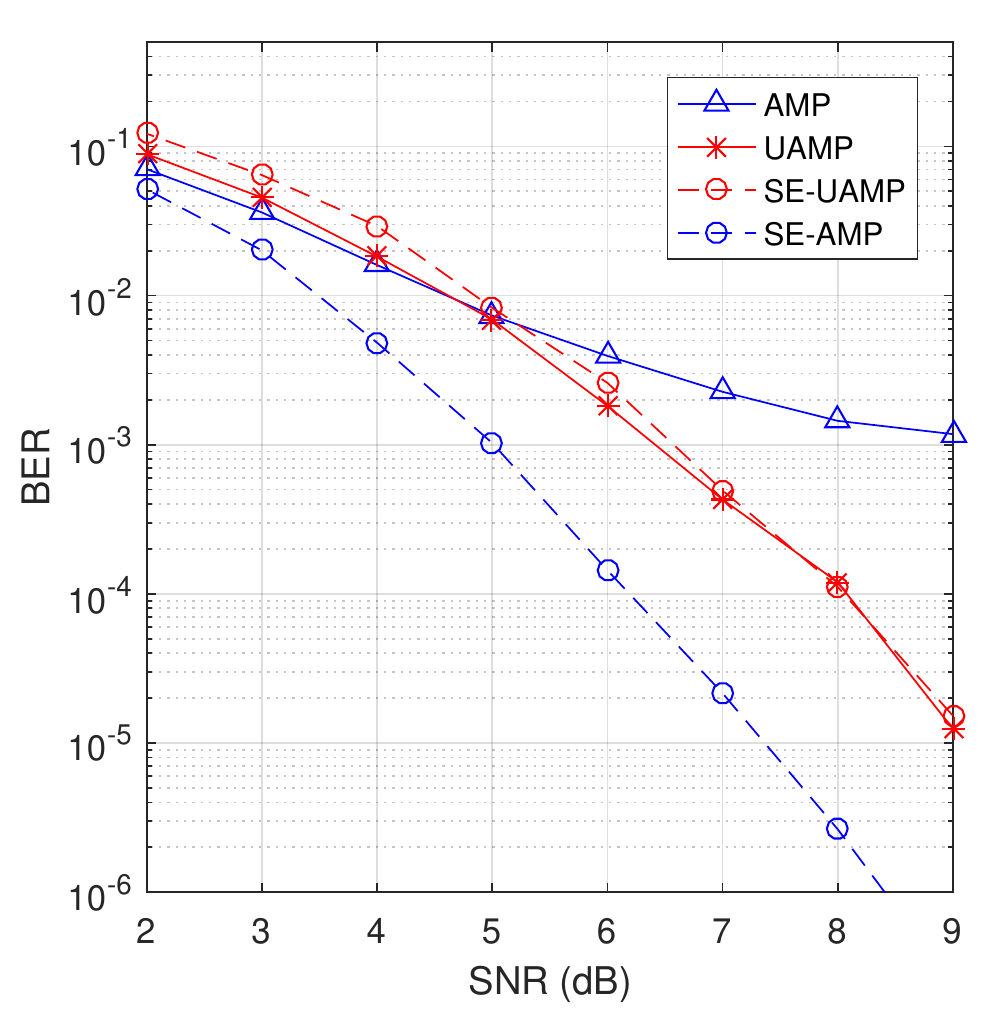}
	\caption {SE and simulated performance of the turbo receiver with the (U)AMP-based detectors, where $P=10$.}
	\label{fig:SEresult}
\end{figure}

We then evaluate the performance of the detectors in a coded OTFS system, where the turbo receiver shown in Fig. \ref{fig:iterative} is employed. We use a rate-1/2 convolutional code with generator $[5,7]_8$, followed by a random interleaver and QPSK modulation. The length of the codeword is $MN$. The BCJR algorithm is used for the SISO decoder. Fig. \ref{fig:RectBerVsSnrCoded} shows the BER performance {of both coded and uncoded systems with} the (U)AMP-based detectors for $P=10$ and $P=14$.
{We can} find that the performance gaps between the AMP-based detector and the UAMP-based detector become larger in the coded system. It can also be seen that the turbo receiver can achieve much better performance (about $3.5-4$dB at the BER of $10^{-4}$) thanks to the iteration between the SISO detector and the SISO decoder.

\begin{figure}[h]
	\centering
	\includegraphics[width=1\columnwidth]{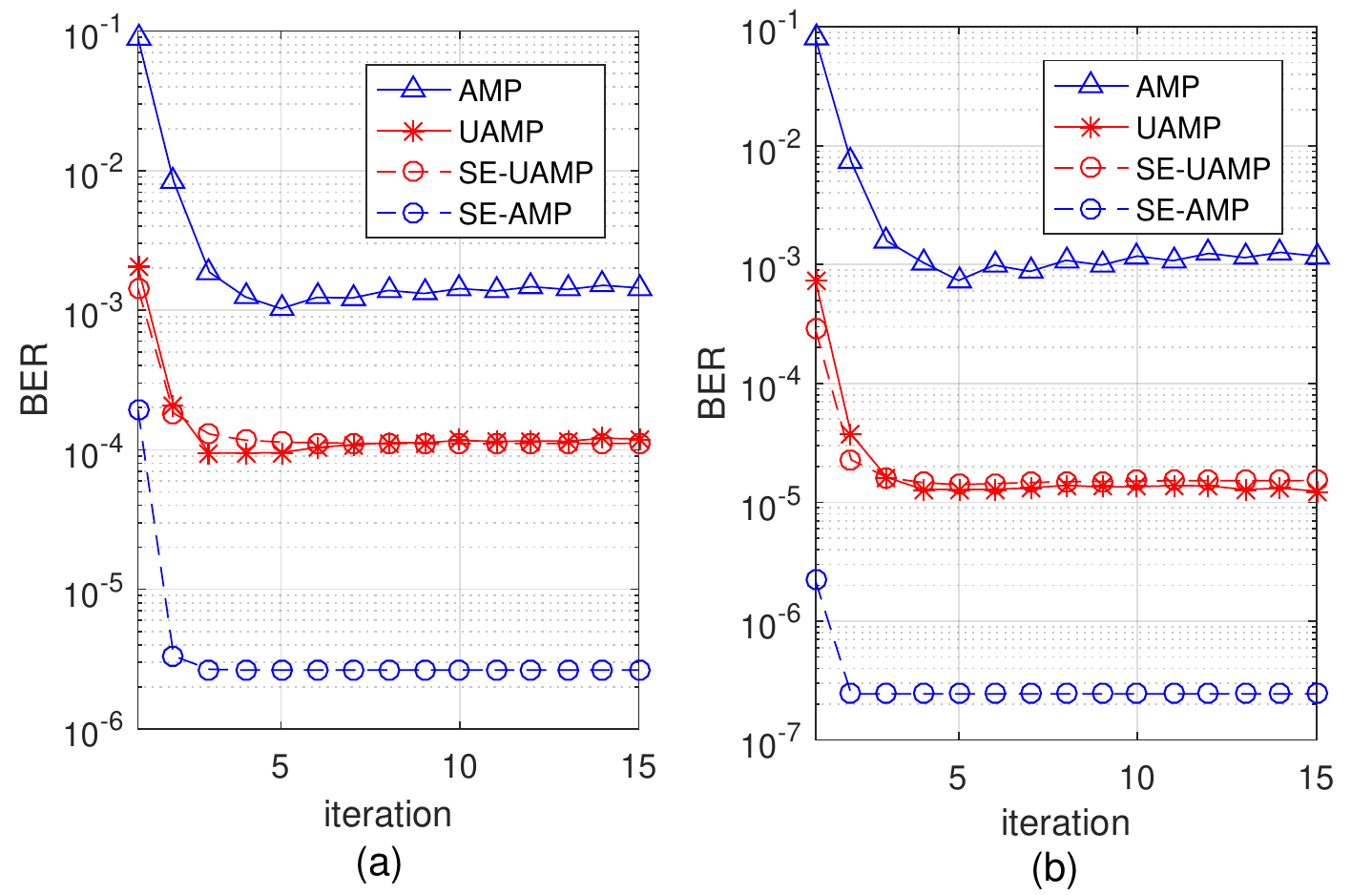}
	\caption {SE and simulated performance of the turbo receiver with the (U)AMP-based detectors (a) SNR = 8dB, and (b) SNR = 9dB.}
	\label{fig:SEresulta}
\end{figure}

Last, we examine the effectiveness of the SE-based BER performance prediction in {Section V}. The coded OTFS system has the same settings as that in Fig. \ref{fig:RectBerVsSnrCoded} (a). The SE performance and the simulated performance versus SNR is shown in Fig. \ref{fig:SEresult}, where we can see that the predicted performance matches well with the simulated performance for the UAMP-based detector. However, for the AMP-based detector, the predicated performance is no longer accurate, and the SE overestimates the performance severely. This is because { the AMP SE is  invalid when the channel matrix deviates from the i.i.d. Gaussian one.} Fig. \ref{fig:SEresulta} shows the predicated performance and simulated performance of the (U)AMP-based receiver versus iteration number at SNR = 8dB and 9dB, where it can be seen that the predicted performance matches well with the simulated performance for the UAMP-based detector.

\section{Conclusions}

In this paper, to address the issues of OFTS detection when the number of channel paths is relatively large and the fractional Doppler shifts have to be considered, we have investigated the design of OTFS detectors based on (U)AMP. In particular, the UAMP-based detectors enjoy the structure of the channel matrix, allowing more efficient implementation while with enhanced performance due to the robustness of UAMP. The investigations have also been extended to joint detection and decoding in a coded OTFS system with a turbo receiver. It has been demonstrated that the UAMP-based detector can outperform the state-of-the-art detectors significantly, and the performance of the system with the UAMP-based detector can be predicted well with the state evolution.

\bibliographystyle{IEEEtran}
\bibliography{IEEEabrv,bibliography}

\begin{thebibliography}{10}
\providecommand{\url}[1]{#1}
\csname url@samestyle\endcsname
\providecommand{\newblock}{\relax}
\providecommand{\bibinfo}[2]{#2}
\providecommand{\BIBentrySTDinterwordspacing}{\spaceskip=0pt\relax}
\providecommand{\BIBentryALTinterwordstretchfactor}{4}
\providecommand{\BIBentryALTinterwordspacing}{\spaceskip=\fontdimen2\font plus
\BIBentryALTinterwordstretchfactor\fontdimen3\font minus
  \fontdimen4\font\relax}
\providecommand{\BIBforeignlanguage}[2]{{%
\expandafter\ifx\csname l@#1\endcsname\relax
\typeout{** WARNING: IEEEtran.bst: No hyphenation pattern has been}%
\typeout{** loaded for the language `#1'. Using the pattern for}%
\typeout{** the default language instead.}%
\else
\language=\csname l@#1\endcsname
\fi
#2}}
\providecommand{\BIBdecl}{\relax}
\BIBdecl

\bibitem{Hadani2017}
R.~Hadani, S.~Rakib, M.~Tsatsanis, A.~Monk, and R.~Calderbank, ``Orthogonal
  time frequency space modulation,'' in \emph{2017 IEEE Wireless Communications
  and Networking Conference (WCNC)}, 2017.

\bibitem{Raviteja2018}
P.~Raviteja, P.~K. T., H.~Yi, and V.~Emanuele, ``Interference cancellation and
  iterative detection for orthogonal time frequency space modulation,''
  \emph{IEEE Transactions on Wireless Communications}, vol.~17, no.~10, pp.
  6501--6515, 2018.

\bibitem{Surabhi2019}
G.~D. {Surabhi}, R.~M. {Augustine}, and A.~{Chockalingam}, ``On the diversity
  of uncoded {OTFS} modulation in doubly-dispersive channels,'' \emph{IEEE
  Transactions on Wireless Communications}, vol.~18, no.~6, pp. 3049--3063,
  June 2019.

\bibitem{OAP}
R.~Hadani and A.~Monk, ``{OTFS}: A new generation of modulation addressing the
  challenges of 5g,'' \emph{ArXiv}, vol. abs/1802.02623, 2018.

\bibitem{shuangyang2020twc}
S.~Li, J.~Yuan, W.~Yuan, Z.~Wei, and D.~W.~K. Ng, ``Performance analysis of
  coded {OTFS} systems over high-mobility channels,'' \emph{IEEE Trans.
  Wireless Commun.}, submitted.

\bibitem{Farhang2018}
A.~{Farhang}, A.~{RezazadehReyhani}, L.~E. {Doyle}, and B.~{Farhang-Boroujeny},
  ``Low complexity modem structure for {OFDM}-based orthogonal time frequency
  space modulation,'' \emph{IEEE Wireless Communications Letters}, vol.~7,
  no.~3, pp. 344--347, June 2018.

\bibitem{LiA2017}
\BIBentryALTinterwordspacing
L.~Li, H.~Wei, Y.~Huang, Y.~Yao, W.~Ling, G.~Chen, P.~Li, and Y.~Cai, ``A
  simple two-stage equalizer with simplified orthogonal time frequency space
  modulation over rapidly time-varying channels,'' \emph{CoRR}, vol.
  abs/1709.02505, 2017. [Online]. Available:
  \url{http://arxiv.org/abs/1709.02505}
\BIBentrySTDinterwordspacing

\bibitem{zemen2017}
\BIBentryALTinterwordspacing
T.~Zemen, M.~Hofer, and D.~Loeschenbrand, ``Low-complexity equalization for
  orthogonal time and frequency signaling ({OTFS}),'' 2017. [Online].
  Available: \url{http://arxiv.org/abs/1710.09916}
\BIBentrySTDinterwordspacing

\bibitem{rakea}
T.~{Thaj} and E.~{Viterbo}, ``Low complexity iterative rake detector for
  orthogonal time frequency space modulation,'' in \emph{2020 IEEE Wireless
  Communications and Networking Conference (WCNC)}, 2020, pp. 1--6.

\bibitem{rake}
\BIBentryALTinterwordspacing
T.~Thaj and E.~Viterbo, ``Low complexity iterative rake decision feedback
  equalizer for zero-padded otfs systems,'' \emph{IEEE Transactions on
  Vehicular Technology}, vol.~69, no.~12, p. 15606–15622, Dec 2020. [Online].
  Available: \url{http://dx.doi.org/10.1109/TVT.2020.3044276}
\BIBentrySTDinterwordspacing

\bibitem{Raviteja2019}
P.~{Raviteja}, E.~{Viterbo}, and Y.~{Hong}, ``{OTFS} performance on static
  multipath channels,'' \emph{IEEE Wireless Communications Letters}, vol.~8,
  no.~3, pp. 745--748, June 2019.

\bibitem{Raviteja2017}
P.~Raviteja, K.~T. Phan, Q.~Jin, Y.~Hong, and E.~Viterbo, ``Low-complexity
  iterative detection for orthogonal time frequency space modulation,''
  \emph{ArXiv}, vol. arxiv.org/abs/1709.09402, 2017.

\bibitem{Yuan2019}
W.~{Yuan}, Z.~{Wei}, J.~{Yuan}, and D.~W.~K. {Ng}, ``A simple variational bayes
  detector for orthogonal time frequency space ({OTFS}) modulation,''
  \emph{IEEE Transactions on Vehicular Technology}, vol.~69, no.~7, pp.
  7976--7980, 2020.

\bibitem{otfslmmserecv}
S.~{Tiwari}, S.~S. {Das}, and V.~{Rangamgari}, ``Low complexity {LMMSE}
  receiver for {OTFS},'' \emph{IEEE Communications Letters}, vol.~23, no.~12,
  pp. 2205--2209, 2019.

\bibitem{Donoho2010a}
D.~L. Donoho, A.~Maleki, and A.~Montanari, ``Message passing algorithms for
  compressed sensing: {I}. motivation and construction,'' in \emph{2010 IEEE
  Information Theory Workshop on Information Theory (ITW 2010, Cairo)}, Jan
  2010, pp. 1--5.

\bibitem{Donoho2010b}
------, ``Message passing algorithms for compressed sensing: {II}. analysis and
  validation,'' in \emph{2010 IEEE Information Theory Workshop on Information
  Theory (ITW 2010, Cairo)}, Jan 2010, pp. 1--5.

\bibitem{Caltagirone2014}
F.~Caltagirone, L.~Zdeborova, and F.~Krzakala, ``On convergence of approximate
  message passing,'' in \emph{2014 IEEE International Symposium on Information
  Theory}, June 2014, pp. 1812--1816.

\bibitem{Guo2013}
Q.~{Guo}, D.~{Huang}, S.~{Nordholm}, J.~{Xi}, and Y.~{Yu}, ``Iterative
  frequency domain equalization with generalized approximate message passing,''
  \emph{IEEE Signal Processing Letters}, vol.~20, no.~6, pp. 559--562, June
  2013.

\bibitem{Guo2015UtAMP}
\BIBentryALTinterwordspacing
Q.~Guo and J.~Xi, ``Approximate message passing with unitary transformation,''
  \emph{CoRR}, vol. abs/1504.04799, 2015. [Online]. Available:
  \url{http://arxiv.org/abs/1504.04799}
\BIBentrySTDinterwordspacing

\bibitem{BiUTAMP}
Z.~{Yuan}, Q.~{Guo}, and M.~{Luo}, ``Approximate message passing with unitary
  transformation for robust bilinear recovery,'' \emph{IEEE Transactions on
  Signal Processing}, vol.~69, pp. 617--630, 2021.

\bibitem{UTAMPSBL}
M.~{Luo}, Q.~{Guo}, D.~{Huang}, and J.~{Xi}, ``Sparse bayesian learning based
  on approximate message passing with unitary transformation,'' in \emph{2019
  IEEE VTS Asia Pacific Wireless Communications Symposium (APWCS)}, 2019, pp.
  1--5.

\bibitem{ISAR}
H.~{Kang}, J.~{Li}, Q.~{Guo}, and M.~{Martorella}, ``Pattern coupled sparse
  bayesian learning based on {UTAMP} for robust high resolution {ISAR}
  imaging,'' \emph{IEEE Sensors Journal}, pp. 1--1, 2020.

\bibitem{DOAUT}
\BIBentryALTinterwordspacing
Y.~Mao, M.~Luo, D.~Gao, and Q.~Guo, ``Low complexity {DOA} estimation using
  {AMP} with unitary transformation and iterative refinement,'' \emph{Digital
  Signal Processing}, p. 102800, 2020. [Online]. Available:
  \url{http://www.sciencedirect.com/science/article/pii/S1051200420301457}
\BIBentrySTDinterwordspacing

\bibitem{Monk2016OTFSO}
\BIBentryALTinterwordspacing
A.~Monk, R.~Hadani, M.~Tsatsanis, and S.~Rakib, ``{OTFS} - orthogonal time
  frequency space,'' \emph{CoRR}, vol. abs/1608.02993, 2016. [Online].
  Available: \url{http://arxiv.org/abs/1608.02993}
\BIBentrySTDinterwordspacing

\bibitem{Raviteja2019Practical}
P.~{Raviteja}, Y.~{Hong}, E.~{Viterbo}, and E.~{Biglieri}, ``Practical
  pulse-shaping waveforms for reduced-cyclic-prefix {OTFS},'' \emph{IEEE
  Transactions on Vehicular Technology}, vol.~68, no.~1, pp. 957--961, 2019.

\bibitem{Rangan2011}
S.~Rangan, ``Generalized approximate message passing for estimation with random
  linear mixing,'' in \emph{Proc. IEEE Int. Symp. on Inform. Theory (ISIT
  2011)}, {Aug.} 2011, pp. 2168--2172.

\bibitem{Tipping}
M.~E. Tipping, ``Sparse bayesian learning and the relevance vector machine,''
  \emph{Journal of Machine Learning Research}, vol.~1, no.~3, pp. 211--244,
  2001.

\bibitem{combine}
E.~{Riegler}, G.~E. {Kirkelund}, C.~N. {Manchon}, M.~{Badiu}, and B.~H.
  {Fleury}, ``Merging belief propagation and the mean field approximation: A
  free energy approach,'' \emph{IEEE Transactions on Information Theory},
  vol.~59, no.~1, pp. 588--602, 2013.

\bibitem{Ahmad2017}
\BIBentryALTinterwordspacing
A.~RezazadehReyhani, A.~Farhang, M.~Ji, R.~Chen, and B.~Farhang-Boroujeny,
  ``Analysis of discrete-time {MIMO OFDM}-based orthogonal time frequency space
  modulation,'' \emph{CoRR}, vol. abs/1710.07900, 2017. [Online]. Available:
  \url{https://arxiv.org/abs/1710.07900}
\BIBentrySTDinterwordspacing

\bibitem{OTFSCP}
A.~{Farhang}, A.~{RezazadehReyhani}, L.~E. {Doyle}, and B.~{Farhang-Boroujeny},
  ``Low complexity modem structure for {OFDM}-based orthogonal time frequency
  space modulation,'' \emph{IEEE Wireless Communications Letters}, vol.~7,
  no.~3, pp. 344--347, 2018.

\bibitem{channelofdm}
M.~{Guillaud} and D.~T.~M. {Slock}, ``Channel modeling and associated
  inter-carrier interference equalization for {OFDM} systems with high doppler
  spread,'' in \emph{2003 IEEE International Conference on Acoustics, Speech,
  and Signal Processing, 2003. Proceedings. (ICASSP '03).}, vol.~4, 2003, pp.
  IV--237.

\bibitem{GuoAConcise}
Q.~{Guo} and D.~D. {Huang}, ``A concise representation for the soft-in soft-out
  {LMMSE} detector,'' \emph{IEEE Communications Letters}, vol.~15, no.~5, pp.
  566--568, 2011.

\bibitem{mmseequa}
M.~{Tuchler}, A.~C. {Singer}, and R.~{Koetter}, ``Minimum mean squared error
  equalization using a priori information,'' \emph{IEEE Transactions on Signal
  Processing}, vol.~50, no.~3, pp. 673--683, 2002.

\bibitem{Guograph}
Q.~{Guo} and L.~{Ping}, ``{LMMSE} turbo equalization based on factor graphs,''
  \emph{IEEE Journal on Selected Areas in Communications}, vol.~26, no.~2, pp.
  311--319, 2008.

\bibitem{turbobook}
B.~Vucetic and J.~Yuan, \emph{Turbo codes: principles and applications}.\hskip
  1em plus 0.5em minus 0.4em\relax Springer Science \& Business Media, 2012,
  vol. 559.

\bibitem{snrevolution}
X.~{Yuan}, Q.~{Guo}, X.~{Wang}, and L.~{Ping}, ``Evolution analysis of low-cost
  iterative equalization in coded linear systems with cyclic prefixes,''
  \emph{IEEE Journal on Selected Areas in Communications}, vol.~26, no.~2, pp.
  301--310, 2008.

\bibitem{Guodouble}
Q.~{Guo}, L.~{Ping}, and D.~{Huang}, ``A low-complexity iterative channel
  estimation and detection technique for doubly selective channels,''
  \emph{IEEE Transactions on Wireless Communications}, vol.~8, no.~8, pp.
  4340--4349, 2009.

\end{thebibliography}

\end{document}